\documentclass[a4paper,12pt]{article} 

\usepackage{tikz}
\usetikzlibrary{calc,trees,positioning,arrows,chains,shapes.geometric,decorations.pathreplacing,decorations.pathmorphing,shapes,matrix,shapes.symbols}

\usepackage{bm}
\usepackage{calligra}
\usepackage{amsmath, amssymb, cite, fancyhdr}
\usepackage{comment}
\usepackage{mathrsfs}
\usepackage{dsfont} %\mathds prettier than \mathbb
\usepackage{graphicx}
\usepackage{hyperref}
\usepackage{amsthm}
\usepackage{bbold}
\usepackage{eufrak}
\usepackage{threeparttable}
\usepackage{scalerel}
\allowdisplaybreaks

\usepackage{titlesec}

\usepackage{parskip}
\hypersetup{
  colorlinks,
  citecolor=red,
  linkcolor=blue,
  linktoc=all
}

\numberwithin{equation}{section}

\usepackage{mathtools}
\usepackage{enumitem}
\DeclareMathAlphabet{\mathcalligra}{T1}{calligra}{m}{n}

\usepackage[nolist,printonlyused,smaller]{acronym} 
\usepackage{xcolor}%enables use of coloured text
\usepackage{etex}
\usepackage{morefloats}

\usepackage{longtable}

\newcommand{\bg}{\begin{equation}}
	\newcommand{\eg}{\end{equation}}
\newcommand{\spl}[1]{\begin{split}#1\end{split}} 
\newcommand{\gl}[1]{\eqref{eq:#1}}
\newcommand{\Gl}[1]{Eq.~\eqref{eq:#1}}

\def\m{_{\mu}}
\def\n{_{\nu}}
\def\M{^{\mu}}
\def\N{^{\nu}}

\def\mn{_{\mu\nu}}
\def\MN{^{\mu\nu}}

\def\rs{_{\rho\sigma}}

\def\RS{^{\rho\sigma}}

\def\MR{^{\mu\rho}}

\def\NR{^{\nu\rho}}

\def\MS{^{\mu\sigma}}

\def\NS{^{\nu\sigma}}

\def\MNRS{^{\mu\nu\rho\sigma}}

\makeatletter
\newcommand*\bigcdot{\cdot}
\newcommand*\bigcdot@[2]{\mathbin{\vcenter{\hbox{\scalebox{#2}{$\m@th#1\bullet$}}}}}
\makeatother

\newcommand{\half}{\frac{1}{2}}
\newcommand{\munu}{{\mu\nu}}

\newcommand{\ssum}[2]{\sum_{#1}^{#2}}

\newcommand{\ita}[1]{\textit{#1}}
\newcommand{\mrm}[1]{\mathrm{#1}}
\newcommand{\lef}{\left}
\newcommand{\ri}{\right}
\newcommand{\e}{\mathrm{e}}

\newcommand{\foh}{\frac{1}{2}}
\newcommand{\Tr}{\operatorname{Tr}}

\newcommand{\D}{\mathrm{d}}

\newcommand{\sgo}{\sqrt{g}}

\newcommand{\sg}[1]{\sqrt{g(#1)}}
\newcommand{\sgbo}{\sqrt{\bar{g}}}

\newcommand{\intD}{\int\!\mathcal{D}}
\newcommand{\dd}{\D^d}

\newcommand{\what}[1]{\widehat{#1}}

\newcommand{\lb}{\lambda}

\newcommand{\inv}{^{-1}}

\newcommand{\bff}{{\bf f}}

\newcommand{\mcM}{{\mathcal{M}}}

\newcommand{\mcO}{{\mathcal{O}}}
\newcommand{\mcP}{{\mathcal{P}}}

\newcommand{\mcS}{{\mathcal{S}}}

\newcommand{\bbC}{{\mathds{C}}}

\newcommand{\bbH}{{\mathds{H}}}

\newcommand{\bbN}{{\mathds{N}}}

\newcommand{\bbP}{{\mathds{P}}}

\newcommand{\bbR}{{\mathds{R}}}
\newcommand{\bbS}{{\mathds{S}}}

\newcommand{\mff}{{\mathfrak{f}}}

\newcommand{\mB}{\mathscr{B}}
\newcommand{\mD}{\mathcal{D}}
\newcommand{\mF}{\mathscr{F}}
\newcommand{\mE}{\mathscr{E}}

\newcommand{\mM}{\mathscr{M}}

\newcommand{\mI}{\mathscr{I}}
\newcommand{\mK}{\mathscr{K}}

\newcommand{\nonum}{\nonumber \\[1.5mm] }
\newcommand{\is }{&\!\!=\!\!&}

\titleformat{\section}[hang]{\normalfont\large\bfseries}{\thesection.}{3mm}{}
\titlespacing{\section}{0mm}{10mm}{2mm}

\titleformat{\subsection}[hang]{\normalfont\bfseries}{\thesubsection}{2mm}{}
\titlespacing{\subsection}{0mm}{8mm}{2mm}

\titleformat{\subsubsection}[hang]{\normalfont\bfseries}{\thesubsubsection}{5mm}{}
\titlespacing{\subsubsection}{0mm}{5mm}{2mm}

\newtheoremstyle{thmstyle}
{7pt} % Space above
{3pt} % Space below
{\itshape} % Body font
{} % Indent amount
{\bfseries} % Theorem head font
{.} % Punctuation after theorem head
{0.5em} % Space after theorem head
{} % Theorem head spec (can be left empty, meaning `normal')

\theoremstyle{remark}{

}

\theoremstyle{thmstyle}{

\newtheorem{theorem}{Theorem}[section]

\newtheorem{lemma}{Lemma}[section]
\newtheorem{corollary}{Corollary}[theorem]
}

\begin{document} 
	\pagestyle{fancy}
	\fancyhf{}
	\renewcommand{\headrulewidth}{0pt}
	\fancyhead[R]{MITP-23-003}
	\cfoot{}
	
	\begin{acronym}
		%\acro{sm}[SM]{Standard Model}
		\acro{qft}[QFT]{quantum field theory}
		\acro{qg}[QG]{quantum gravity}
		\acro{gr}[GR]{general relativity}
		\acro{rhs}[RHS]{right-hand side}
		\acro{lhs}[LHS]{left-hand side}
		\acro{rs}[RS]{rigid spacetime}
		\acro{sc}[SC]{selfconsistent}
		\acro{ol}[1L]{one loop}
		\acro{ea}[EA]{effective action}
		%\acro{eaa}[EAA]{effective average action}
		\acro{uv}[UV]{ultraviolet}
		\acro{ir}[IR]{infrared}
		\acro{brst}[BRST]{Becchi-Rouet-Stora-Tyutin}
		%\acro{frge}[FRGE]{functional renormalization group equation}
		\acro{rg}[RG]{renormalization group}
		\acro{frg}[FRG]{functional renormalization group}
		\acro{set}[SET]{stress-energy tensor}
	\end{acronym}

{\hypersetup{linkcolor=black}
	\begin{center}
		{\bf\Large $N$-cutoff regularization for fields on hyperbolic space}
		\vspace{2cm}

		{\sc Rudrajit~Banerjee}\footnote{email: \ttfamily\href{mailto:rudrajit.banerjee@oist.jp}{rudrajit.banerjee@oist.jp}},
		{\sc Maximilian~Becker}\footnote{email: \ttfamily\href{mailto:M.Becker@hef.ru.nl}{M.Becker@hef.ru.nl}},
		{\sc and~Renata Ferrero}\footnote{email: \ttfamily\href{mailto:rferrero@uni-mainz.de}{rferrero@uni-mainz.de}}
		\\[8mm]
		
		{\small\sl $^1$Okinawa Institute of Science and Technology Graduate University,}\\
		{\small\sl 1919-1, Tancha, Onna, Kunigami District, Okinawa 904-0495, Japan}
		\\[5mm]
		{\small\sl $^2$Institute for Mathematics, Astrophysics and Particle Physics (IMAPP),}\\
		{\small \sl Radboud University, Heyendaalseweg 135, 6525 AJ Nijmegen, The Netherlands}
		\\[5mm]
		{\small \sl $^3$Institute of Physics (THEP), University of Mainz,}\\
		{\small \sl Staudingerweg 7, D-55128 Mainz, Germany}
		\vspace{18mm}
		
	\end{center}
}

\begin{abstract}
We apply a novel background independent regularization scheme, the {\itshape $N$-cutoffs}, to self-consistently quantize scalar and metric fluctuations on the maximally symmetric but non-compact hyperbolic space. For quantum matter fields on a classical background or full Quantum Einstein Gravity (regarded here as an effective field theory) treated in the background field formalism, the $N$-cutoff is an ultraviolet regularization of the fields' mode content that is independent of the background hyperbolic space metric. For each $N>0$, the regularized system backreacts on the geometry to dynamically determine the self-consistent background metric. The limit in which the regularization is removed then automatically yields the `physically correct' spacetime on which the resulting quantum field theory lives. When self-consistently quantized with the $N$-cutoff, we find that without any fine-tuning of parameters, the vacuum fluctuations of scalar and (linearized) graviton fields do not lead to the usual cosmological constant problem of a curvature singularity. Instead, the presence of increasingly many field modes tends to reduce the negative curvature of hyperbolic space, leading to vanishing values in the limit of removing the cutoff.

	%We apply a novel background-independent and scale-free quantization and regularization scheme, the {\itshape $N$-cutoffs}, to scalar fields and metric fluctuations on the maximally symmetric but non-compact  hyperbolic space. The $N$-cutoffs are a regularization on the spectrum of the fields' fluctuation modes, and are implemented on a continuous quantum number here. For both the scalar field and metric fluctuations, we find that the inclusion of increasingly many field modes tends  to reduce the negative curvature of the hyperbolic space, leading to vanishing values in the limit of removing the cutoff. This phenomenon is analogous to the results found for those fields on a self-consistent spherical background.   
\end{abstract}

\newpage

\pagestyle{fancy}
\fancyhf{}
\fancyhead[R]{}
\renewcommand{\headrulewidth}{0pt}
\cfoot{}

\begin{samepage}
	{\hypersetup{linkcolor=black}
		\tableofcontents}
\end{samepage}

\newpage

\setcounter{page}{1}
\cfoot{\thepage}

\section{Introduction}
\label{sec:Int}

Of the known fundamental interactions, gravity carries the unique distinction of determining the very stage on which these interactions manifest -- the geometry of spacetime itself. In this spirit, any background structures appearing in a quantum theory must be self-consistently determined as a prediction of the theory, rather than being fixed a priori. When formulating a \ac{qft} on a curved background, this requirement of self-consistency may be imposed at either of the two (logically independent) steps necessary for defining the theory -- 
regularization or renormalization. Imposition of self-consistency after the latter stage is standard:
 treating spacetime purely classically, this accounts for the backreaction of the non-gravitational quantum fields in semi-classical gravity \cite{Kie-Why, Padmanabhan:1988ur, Birrell:1982ix, Parker:2009uva}, while in full quantum gravity this leads to `quantum spacetime' emerging as a prediction of the theory \cite{Reuter:2019byg}. Alternatively, demanding self-consistency already at the level of the regularized theory potentially leads to \ac{qft}s  distinct from the standard approach when the regulator is removed.  This novel approach has recently been developed in \cite{Max1,Max2}, and is explored on a technical level through a regularization procedure dubbed the `$N$-type cutoffs' or `$N$-cutoffs' for short.

$N$-cutoffs are a technical tool for metric-independent \ac{uv} regularization of a (bare) QFT so that a self-consistent determination of any background structures can be implemented already at the level of the bare regularized theory. In brief, one considers a sequence of {\itshape gravity-coupled approximants}, i.e. quantum systems coupled to a background metric, indexed by a dimensionless number $N$ (which may be continuous or discrete) that designates `how many' quantum modes are allowed in the path integral, thereby functioning as a UV regulator. Importantly, the choice of $N$ is independent of the background metric. Performing the path integral for each approximant, varying the induced gravitational action yields the  `quantum equations of motion' (or effective Einstein equations), which may be solved to obtain the self-consistent background metric for each $N$.   Finally, taking the $N\to \infty$ limit yields the self-consistent spacetime that the concomitant QFT lives on.

For the case of {\itshape compact} manifolds, the  consequences of this `self-consistent quantization' program were explored for the paradigmatic setting of spherical geometries in 
\cite{Max1,Max2}. 
 For both scalar and metric fluctuations, contrary to the naive expectation that the presence of increasingly many quantum modes leads to a diverging (bare) vacuum energy density and hence a degenerate spacetime (without sufficient fine-tuning) \cite{Weinberg:1988cp,Carroll:2000fy,Straumann:2002tv, martincc}, the above self-consistency condition yields the striking result that the curvature of the sphere for the resulting sequence of gravity coupled approximants {\itshape vanishes} in the $N\to \infty$  (i.e. QFT) limit.\footnote{This procedure is similar in spirit to the implementation of a (background independent) cutoff used for fuzzy spaces \cite{Madore:1991bw, MADORE1991245, Fiore:2022twy, FIORE2018423}.}
Crucial for this  amelioration of the ``cosmological constant problem'' is the feature that the cutoff procedure, determined by the dimensionless number $N$, is independent of the background metric. In contradistinction, the oft-used momentum cutoff, where the background metric enters in determining the momentum-scale at which modes are restricted,  leads to the usual vacuum energy divergences. 
Moreover, by continuing the sphere to de Sitter space, it was shown the $N$-cutoff regularized systems may be interpreted (for finite $N$) as comprising those microscopic degrees of freedom that are counted in the Bekenstein-Hawking entropy.

In this paper we initiate a study of the $N$-cutoff formalism for self-consistent quantization on {\itshape non-compact} Riemannian manifolds, focussing on the prototypical example of hyperbolic space. Indeed, hyperbolic geometry is frequently employed in studies of quantum gravity and quantum field theory on curved backgrounds, see e.g. \cite{Benedetti:2014gja, fallshyp, ferrarihyp, gfthyp}.
 Although hyperbolic space  is maximally symmetric, the extension of the $N$-cutoff program from the sphere is less straightforward than it may seem. For example, scalar \ac{qft}s on hyperbolic space have markedly distinct behaviors to their spherical counterparts \cite{Benedetti:2014gja}. Likewise, mathematical subtleties due to  non-compactness need to be accounted for when formulating the gravity coupled approximants. This paper serves as a proof of concept that the $N$-cutoff program has potentially interesting physical consequences on non-compact spacetimes.  In overview, both scalar and metric fluctuations yield qualitatively similar self-consistent geometries, and our main results are:
 \begin{enumerate}[leftmargin=6mm, rightmargin=-0mm, label=(\roman*)]
 
  %\red{Add comment that this has a $N=0$ (i.e. classical limit), but does not have a QFT limit... construction seems to prefer positive c.c. for QFT limit?}
  \item Remarkably, for a non-negative bare cosmological constant, one finds a  self-consistent hyperbolic space solution  (with constant negative curvature) for all $N>0$, with the striking feature that the resulting curvature  {\itshape vanishes} as $N\to \infty$. Since the classical Einstein equations (i.e. for $N=0$) possess no hyperbolic space solutions in this case, this sequence of self-consistent manifolds owes its existence entirely to quantum effects.
   \item For a negative bare cosmological constant, hyperbolic space can only self-consistently support quantum modes up to a finite $N_{\rm max}$, beyond which there are no self-consistent solutions. 
\end{enumerate}

\medskip

The structure of this paper is as follows. In Section \ref{sec:Ncutoff} we present the general features of the `self-consistent quantization via $N$-cutoffs' program, and subsequently construct the $N$-cutoff regularized approximants on $d$-dimensional hyperbolic space. In Section \ref{sec:fields} we briefly review the (formal) path integral quantization of scalar and metric fluctuations on a Riemannian manifold, obtaining the induced gravitational one-loop effective action and associated \ac{set} in each case. 
%Next, in Section \ref{sechyp} we construct the $N$-cutoffs on the $d$-dimensional hyperboloid $\bbH^d(L)$, and apply these to the results of the preceeding section to obtain the regularized one-loop effective action $\Gamma_{\rm 1L}[g]_N$ and \ac{set} $T_\munu[g]_N$. 
Next,  the resulting `self-consistent $N$-geometries' are obtained as solutions of the equations of motion corresponding to the gravity coupled approximants. 
 Finally, we present our conclusions and an outlook in Section~\ref{sec:conclusion}.

%%%%%%%%%%%%%%%%%%%
%	New section 
%%%%%%%%%%%%%%%%%%%

\section{$N$-cutoffs and the hyperbolic space}
\label{sec:Ncutoff}

\subsection{$N$-cutoffs}
\label{ssec:2_1}

\noindent
Recently, a regularization procedure for matter fields in contact with classical gravity, called $N$-type cutoffs (or $N$-cutoffs for short) has been advocated for in~\cite{Max1,Max2}. Conceptually, this regularization scheme is based on three requirements in a logical sequence:
\begin{align*}
	&\textbf{(R1)}\ \text{Background independence,}\\
&\textbf{(R2)}\ \text{Gravity-coupled approximants,}\\
&\textbf{(R3)}\ \text{$N$-type cutoffs.}
\end{align*}
The first requirement \textbf{(R1)} of background independence enforces the entire background spacetime geometry, on which the matter fields `live', to be determined by dynamical principles.\footnote{While a (mathematically) rigorous definition of the term `background independence' has not yet been formulated (c.f. \cite{Giu-BI}), one may regard the requirement of {\itshape all background structures being dynamical} as our working definition of `background independence'.} We refer to imposing these dynamical principles on the background geometry as imposing the self-consistency condition. This self-consistency condition is of particular importance for applying $N$-cutoffs to continuum-based quantum gravity, which heavily relies on the background-field technique.

The second requirement \textbf{(R2)} now states that the self-consistency condition must be invoked already at the level of the \ac{uv} regularized \ac{qft}. Thereby the  regularized system, called an `approximant $\mathsf{App}(\mathbf f)$' (where $\bff$ parameterizes the degrees of freedom allowed by the \ac{uv} regularization), shall be interpretable as a physically realizable system in its own right. The quantum modes comprising   $\mathsf{App}(\mathbf f)$ subsequently backreact on the classical metric of the background spacetime, on which the \ac{qft} is being formulated, via the \ac{set} obtained from $\mathsf{App}(\mathbf f)$. This determines a family  $(\mathsf{App}(\mathbf f),g\mn^\mrm{sc})_\bff$  of approximant systems coupled to a self-consistently determined metric tensor, parameterized by $\bff$. Given that each element of this sequence represents a physically realizable system, one may hope for more meaningful \ac{qft}s in the limit `$\mathbf f\to\infty$' (i.e. when {\itshape all} degrees of freedom are included). Additionally, this construction also encompasses the peculiar possibility that  physical reality is better described by $(\mathsf{App}(\mathbf f),g\mn^\mrm{sc})$ at some finite $\mathbf f$. This possibility can be viewed as a way of reaching observable discreteness of spacetime at high energies from a continuum-based quantum gravity approach.

%The second requirement \textbf{(R2)} now states that the self-consistency condition must be invoked already at the level of the regularized \ac{qft}. Thereby, the regularized \ac{qft} is required to constitute a quantum mechanical system with $\mathbf f<\infty$ degrees of freedom that shall be interpretable as a physically realizable system in its own right. The regularized system then is called an approximant, $\mathsf{App}(\mathbf f)$, and the regularized matter field's $\mathbf f$ degrees of freedom subsequently backreact to the classical metric of the background spacetime, on which the \ac{qft} is formulated, via the \ac{set} obtained from $\mathsf{App}(\mathbf f)$. As a result, one obtains a sequence $(\mathsf{App}(\mathbf f),g\mn^\mrm{sc})_{\mathbf f=0,1,2,\dots}$ of approximant systems together with a self-consistent metric tensor. Given that each element of this sequence represents a physically realizable system, the hope here is to find more meaningful \ac{qft}s in the limit $\mathbf f\to\infty$. Additionally, this construction also encompasses the peculiar possibility that our physical reality might be better described by $(\mathsf{App}(\mathbf f),g\mn^\mrm{sc})$ at some finite $\mathbf f$. This possibility can for instance be viewed as a way of reaching observable discreteness of spacetime at high energies from a continuum-based quantum gravity approach.

The third requirement \textbf{(R3)} is a  technical realization of \textbf{(R1)} and \textbf{(R2)} via a {\itshape metric independent} \ac{uv} regularization procedure called the $N$-cutoff, introduced for compact manifolds in \cite{Max1}. To formulate this, consider the path integral quantization of a field $\phi$
 on a background $(\mcM, g)$ belonging to some collection of compact Riemannian manifolds from which the self-consistent metrics are to be determined. 
The formal (Euclidean) path integral  on $(\mcM, g)$
\bg
\label{eq:PI}
Z_g=\int_{\mF_g}\mD_g\phi\,\e^{-S_g[\phi]}\,,
\eg
with the subscript `$g$' denoting the metric dependence, is \ac{uv} regularized as follows. On each background, the function space $\mF_g$ to which the field $\phi$ belongs (typically an $L^2$-space) has a countable orthonormal basis of functions
\bg
\label{eq:basis}
\mB_g=\big\{w_n[g]\,\big|\,n\in \bbN_0\big\} \, ,
\eg
where the selection of this basis is equivalent to the determination of what the individual quantum modes are. Then, one can expand $\phi(x)=\sum_n\phi_n w_n[g](x)$ and the measure will factorize as $\mD\phi\propto\prod_{n\in \bbN_0}\D\phi_n$, thus turning the path integral \gl{PI} into an integral over the field's modes $\phi_n$.

At the heart of the $N$-cutoff construction  is a {\itshape metric independent} way of `counting modes' so as to determine how many modes are allowed in the regularized system. This may be formalized by introducing a sequence of subsets $I_N\subseteq \bbN_0$ of the index set, labeled by a dimensionless $N\geq 0$, such that $I_0=\emptyset,\,I_{N_1}\subseteq I_{N_2}$ when $N_1\leq N_2$, and $\bigcup_{N}I_N=\bbN_0$. This yields a family of regularized path integrals 
\bg
\label{eq:PIN}
Z_{g,N}=\int_{\mF_{g,N}}\mD_g\phi\,\e^{-S_g[\phi]} \, ,
\eg
where  the integration space  $\mF_{g,N}$ is spanned by the sub-base $\mB_{g,N}:=\big\{w_n[g]\,\big|\,n\in I_N\big\}$.
 The degrees of freedom of the regularized system then also become a function of $N$, i.e., $\mathbf f_\mrm{reg}=\mathbf f(N)$. The main advantage of this construction is that the degrees of freedom comprising each regularized system are determined independently of the background metric. Indeed, while the individual quantum modes $w_n[g]$ and integration spaces $\mF_{g,N}$ are metric dependent, whether a given quantum mode is included in the degrees of freedom at some $N$ is determined at the level of the index set $I_N$, where no metric enters.  As a result, the background metric may remain fully dynamical in the regularization procedure, complying with the requirement \textbf{(R2)}.

The selection of the function basis \gl{basis} is constrained by the requirement that for each $N$ the regularized system on the manifold $(\mcM,g)$ -- and therewith the basis $\mB_{g,N}$ -- carries a representation of the background's symmetry group. Otherwise, the self-consistency condition would be ill-posed as the SET arising from the regularization procedure would not carry the same symmetries as the \ac{lhs} of the equations of motion. 
A possible way of constructing sub-bases $\mB_{g,N}$ which meet the symmetry condition is via the eigenfuctions of (metric-dependent) operators, such as the Laplacian $-\Box_g$,\footnote{We adopt the labeling convention that $n'\geq n\implies \lb_{n'}[g]\geq \lb_n[g]$.}
\bg
\label{eq:EV}
-\Box_g u_n[g](x)=\lambda_n[g] u_n[g](x) \, ,
\eg
whose eigenfunctions comprise a countable orthonormal basis on a compact Riemannian manifold.
% The ``adiabatic'' metric-dependence of the eigenfunctions $u_n[g]$ does not come amiss here because it has no influence on the decision whether a field mode is integrated out or not -- it is a mere additional metric-dependence that will enter the self-consistency condition. 
 For the prototypical case of a compact Riemannian manifold, the $d$-sphere $\bbS^d(L)$ with radius $L$, $N$-cutoffs have been constructed in this way \cite{Max1,Max2}. 

\subsection{$N$-cutoffs for scalar and metric fluctuations on  hyperbolic space}
\label{ssec:2_2}

\noindent
Turning to non-compact Riemannian manifolds, as a proof of concept we focus on the prototypical setting of hyperbolic space. In this section, we construct the $N$-cutoffs for scalar, vector, and tensor modes\footnote{We regard the present work as a proof-of-concept purely for the gravitational case. This requires the study of scalar, vector, and tensor modes. Nevertheless, we expect the $N$-cutoff regularization scheme to extend straightforwardly also to the fermionic case. As usual, we expect the fermionic modes to contribute with the opposite sign of the bosonic field species (c.f. \cite{Reitz:2023ezz}).} on $d$-dimensional hyperbolic space with radius $L$ (see below). As emphasized in Section \ref{sec:Ncutoff}, at the heart of this construction is the identification of metric independent way of `counting modes' on hyperbolic space.\footnote{For generic non-compact Riemannian manifolds, the separability of the associated $L^2$ spaces guarantees the existence of countable orthonormal bases, through which the $N$-cutoffs may be constructed analogously. We defer the details to a separate publication.} 

The model of hyperbolic space studied here is the  hyperboloid $\bbH^d(L)$ of dimension $d$ and radius $L$, which is the non-compact submanifold of $(d+1)$-dimensional Minkowski space $\mathds{R}^{1,d}$,  defined by
\begin{equation}\label{eq:E1}
	\bbH^d(L) = \left\{x \in \mathds{R}^{1,d}\;\big|\;(x^0)^2-(x^1)^2-\ldots -(x^d)^2 = L^2\,,\, x^0>0\right\}\,.
\end{equation}
The hyperboloid $\bbH^d(L)$, together with the metric it inherits as a submanifold, is a maximally symmetric Riemannian manifold.
 Its metric is given by the line element (in standard coordinates)
\begin{eqnarray}\label{eq:E2}
	\text{d}s^2 \is L^2\big(\text{d}y^2 + \sinh^2 (y) \text{d}\Omega_{d-1}^2 \big)
\end{eqnarray}
with  $\text{d}\Omega_{d-1}^2$ the line element of the unit $(d-1)$-sphere $\bbS^{d-1}(1) \equiv\bbS^{d-1}$. The line element  \gl{E2} may be written in the from   $g\mn(x)=L^2\gamma\mn(x)$, with $\gamma\mn$ the metric of the unit hyperboloid $\bbH^d(1)$, due to maximal symmetry of $\bbH^d(L)$.

In order to determine the self-consistent backgrounds corresponding to scalar and metric fluctuations on the family of hyperboloids $(\bbH^d(L),g(L))_{L\in (0,\infty)}$, the $N$-cutoffs are constructed  for  scalar ($\mrm S$), vector ($\mrm V$) and symmetric rank-2 tensor fields ($\mrm{ST}$). Further, the function space basis $\mB_g$, and associated  sub-bases  $\mB_{g,N}$, are constructed from the eigenfunctions of the Laplacian $-\Box_g$ to ensure that the regularized approximant $\mathsf{App}(\mathbf f(N))$ carries a representation of the hyperboloid's symmetry group $SO(1,d)$.

The Laplacian $-\Box_g$ acting on the above field species' has a fully continuous spectrum (i.e., there is no discrete part). Writing the eigenvalue problem as
\begin{eqnarray}\label{eq:E4}
	-	\Box_g u_{\lambda l}^\text{I}(x) = \mathscr{E}_\lambda ^\text{I} u_{\lambda l  }^\text{I}(x)
\end{eqnarray}
where $\mrm I=\mrm S; \mrm T, \mrm L; \mrm{TT}, \mrm{LTT}, \mrm{LLT}, \mrm{tr}$ is an index labeling the type of field under consideration\footnote{Here, we have decomposed vector fields (V) into their transverse (T) and longitudinal (L) part, and symmetric rank-2 tensor fields (ST) into their transverse-traceless (TT), longitudinal-transverse-traceless (LTT), longitudinal-longitudinal-traceless (LLT) and trace (tr) part. For details we refer the reader to appendix~\ref{sec:deffields}.}, $\mathscr{E}_\lambda ^\text{I}$ is the eigenvalue depending on the continuous label $\lambda \in [0, \infty)=: \bbR_{\geq 0}$, $l\in \mI(\rm I)$ is a multi-index labeling the degeneracy of the eigenvalue $\mathscr{E}_\lambda^\text{I}$, and $u_{\lambda l}^\text{I}$ is an eigenfunction\footnote{With the spectrum being purely continuous, these eigenfunctions do not live in the corresponding space of square-integrable fields on $\bbH^d(L)$. Rather, these generalized eigenvectors live in an associated Rigged Hilbert space.}  associated to the eigenvalue $\mathscr{E}_\lambda^\text{I}$.
The spectral values $\mathscr{E}^\text{I}_\lb$  and associated eigenfunctions $u_{\lb l}^\text{I}$ (with $\lb\in \bbR_{\geq 0}$) on $\bbH^d(L)$ may be obtained from those of the sphere through analytic continuation \cite{hyp1, Benedetti:2014gja, Demmel:2014sga}. In more detail, on the $d$-sphere $\bbS^d(L)$  the eigenvalue problem for field species I reads 
\begin{eqnarray}\label{eq:E5b}
	-	\Box_g u_{n l}^\text{I}(x) = \mathscr{E}_n ^\text{I} u_{n l  }^\text{I}(x)\,,\quad n\in \bbN_0\,.
\end{eqnarray}
Using standard polar coordinates in which the line element on  $\bbS^d(L)$ is $\D s^2=L^2 \D\chi^2+L^2 \sin^2(\chi) \D\Omega_{d-1}^2$, the eigenfunctions and associated eigenvalues on $\bbH^d(L)$ may with hindsight be obtained through the analytic continuation,
\begin{eqnarray}\label{eq:E6}
	\chi = iy\,,\quad n=-\rho+i\lb \,,\quad  \rho = \frac{d-1}{2}\,,
\end{eqnarray}
resulting in the practical formula 
\begin{equation}\label{eq:E7}
\left(\mathscr{E}_\lambda^\text{I}\right)_{\bbH^d(L)} = -\left(\mathscr{E}_n^\text{I}\right)_{\bbS^d(L)} \Bigg|_{n = -\rho + i \lambda}\,,\quad \lb \in \bbR_{\geq 0}\,.
\end{equation}
The set of eigenfunctions $\left\{u_{\lambda l}^\text{I}\right\}$ obtained in this way thus forms an  orthonormal and complete basis for the Hilbert space of square integral fields of species I on $\bbH^d(L)$. 
The eigenfunctions for $\mrm I=\mrm S,\mrm T,{\rm TT}$ are well known \cite{Camporesi:1990wm, hyp1, Camporesi:1995fb}, and for each $\lb\in \bbR_{\geq 0}$ the set $\{u^\text{I}_{\lb l}\}_{l\in \mI(\rm I)}$  may be regarded as furnishing a basis for a unitary irreducible representation of the action of $SO(1,d)$ on $\bbH^d(L)$.  The remaining eigenfunctions can be constructed from these via the isomorphisms~\gl{C3}, \gl{C11}, \gl{C13}, and~\gl{C14},
\bg\label{eq:E13}
\spl{
&(u^\text{L}_{\lambda l})\m (x)=\lef(P^{\mrm L,\mrm S}\ri)\m u^\text{S}_{\lambda l} (x)\, ,\\
&(u^\text{LTT}_{\lambda l})\mn (x)=\lef(P^{\mrm{LTT,}\mrm{T}}\ri)\n (u^\text{T}_{\lambda l})\m (x)\, ,\\
&(u^\text{LLT}_{\lambda l})\mn (x)=\lef(P^{\mrm{LLT,}\mrm{S}}\ri)\mn u^\text{S}_{\lambda l} (x)\,,\\
&(u^\text{tr}_{\lambda l})\mn (x)=\lef(P^{\mrm{tr,}\mrm{S}}\ri)\mn u^\text{S}_{\lambda l} (x)\,.
}
\eg

The key observation for constructing the $N$-cutoff regularization is that for the entire family of hyperboloids $(\bbH^d(L),g(L))_{L\in (0,\infty)}$, the eigenvalues for each field species I are of the form $\mathscr{E}_\lambda^\text{I}(L)=L^{-2}f_{\rm I}(\lb)$.\footnote{See Table \ref{table:eigenvalues} for the explicit form of the functions $f_{\rm I}(\lb)$.} Accordingly, the allowed degrees of freedom may be parameterized by the dimensionless label $\lb\in \bbR_{\geq 0}$, {\itshape independently of $L$} (i.e. independent of the metric). Thus, the sub-bases $\mB_{g,N}$   are constituted by those  eigenfunctions in \Gl{E4} with the maximal value of the eigenvalue $\lambda$ truncated at the dimensionless \ac{uv}-cutoff parameter $N\in[0,\infty)$, yielding the regularized  function spaces
\bg\label{eq:F5}
\spl{
&L_{\mrm{S}_N}^2\lef(\bbH^d(L),L^2\gamma\mn\ri)=\mrm{span}\,\lef\{u_{\lambda l}^{\mrm{S}}\ \big|\ \lambda\in[0,N]\ ,\ l\in\mI(S)\ri\} \, ,\\[3mm]
&L_{\mrm{V}_N}^2\lef(\bbH^d(L),L^2\gamma\mn\ri)=\bigoplus_{\mrm I=\mrm T,\mrm L}\mrm{span}\,\lef\{u_{\lambda l}^{\mrm{I}}\ \big|\ \lambda\in[0,N]\ ,\ l\in\mI({\rm I})\ri\} \, ,\\[2mm]
&L_{\mrm{ST}_N}^2\lef(\bbH^d(L),L^2\gamma\mn\ri)=\bigoplus_{\mrm I=\mrm{TT},\mrm{LTT},\mrm{LLT},\mrm{tr}}\mrm{span}\,\lef\{u_{\lambda l}^{\mrm{I}}\ \big|\ \lambda\in[0,N]\ ,\ l\in\mI({\rm I})\ri\} \, ,
}
\eg
where $\mrm S_N$, $\mrm V_N$, and $\mrm{ST}_N$ denote the $N$-cutoff regularized field species.

%%%%%%%%%%%%%%%%%%%
%	New subsection 
%%%%%%%%%%%%%%%%%%%

\subsection{$N$-cutoff regularized operator traces}
\label{ssec:2_3}

In Section \ref{ssec:set}, the resulting self-consistency equations \gl{D25} and \gl{D26} are to be expressed in terms of $N$-cutoff regularized operator traces of functions of $-\Box_g$, c.f. Eqs. \gl{D29b}, \gl{D29c}. 
An essential ingredient for evaluating these (regularized) traces is the  spectral function for field species I (also known as the Plancherel measure) \cite{hyp1},
\bg\label{eq:E16}
\mu^\text{I}(\lambda)\!:=\! \frac{\pi \Omega_{d-1}L^d}{2^{d-2}}\frac{1}{\dim(\mrm{I})}\sum_{l\in\mI({\rm I})} u^\text{I}_{\lambda l} (x_0)^*\cdot\,u^\text{I}_{\lambda l} (x_0) \, ,
\eg
which encodes information about the multiplicity of the eigenvalues $\mE_\lambda^\mrm{I}$ of $-\Box_g$.
Here, $x_0$ is arbitrarily chosen point on $\bbH^d(L)$ (which may be thought of as the `origin'), the dot `$\,\cdot\,$' denotes the contraction of the eigenfunctions' tensor structure, and
\bg\label{eq:E17}
\dim(\mrm I)=\begin{cases}
	& 1\quad \text{for}\ \mrm I=\mrm S,\mrm L,\mrm{LLT},\mrm{tr}\\[2mm]
	&(d-1)\quad \text{for}\ \mrm I=\mrm{T},\mrm{LTT}\\[2mm]
	&(d+1)(d-2)/2 \quad\text{for}\ \mrm I=\mrm{TT} 
\end{cases}\,.
\eg
It is to be emphasized that the \ac{rhs} of definition \gl{E16} is {\itshape independent} of the choice of $x_0$. Indeed it can be shown that the functions
\begin{subequations}
\begin{align}\label{E18a}
	d_\lambda ^\text{S}(x) &:=  \sum_{l\in\mI(\mrm{S})}\left(u^\text{S}_{\lambda l}\right)^* (x)\cdot\,\left(u^\text{S}_{\lambda l}\right) (x) \, ,
	\\ \label{E18b}
	d_\lambda ^\text{T}(x) &:=   \sum_{l\in\mI(\mrm{T})} g^{\mu \nu}(x)\left(u^\text{T}_{\lambda l}\right)_\mu^* (x)\cdot\,\left(u^\text{T}_{\lambda l}\right)_\nu (x) \, ,
	\\ \label{E18c}
	d_\lambda ^\text{TT}(x) &:= \sum_{l\in\mI(\mrm{TT})}g^{\mu \alpha}(x)g^{\nu \beta}(x)\left(u^\text{TT}_{\lambda l}\right)_{\alpha \beta}^*(x)\cdot\,\left(u^\text{TT}_{\lambda l}\right)_{\mu \nu}(x)\,,
\end{align}
\end{subequations}
are constant in $x$ on $\bbH^d(L)$, i.e. $d_\lb^\text{I}(x)\equiv d_\lb^\text{I}$. These relations are the hyperbolic space analogues of the spherical harmonic addition theorems, and
follow from the fact that for each field species I and label $\lb\in \bbR_{\geq 0}$, the eigenfunctions $u^{\rm I}_{\lb l}$ carry an irreducible representation of $SO(1,d)$, which acts transitively on the $\bbH^d(L)$. We remark that the derivation of the relations \eqref{E18a}-\eqref{E18c} is more subtle as compared to the sphere, since the non-compactness of hyperbolic space entails that the summations are over infinitely many terms; a proof for the scalar case, together with a discussion of the associated subtleties, is presented in Appendix \ref{sec:maths}.\footnote{We expect that the techniques of proof for the scalar case generalize to higher spin fields, with potentially some additional technical input due to the tensor structure.} 

From the definition of the eigenfunctions for the species ${\rm I}=\text{L, LTT, LLT, tr}$, Eqs.~\gl{E13}, together with the relations \eqref{E18a}-\eqref{E18c}, it follows that
\bg\label{eq:E19}
\mu^\mrm{S}(\lambda)\equiv\mu^\mrm{L}(\lambda)\equiv \mu^\mrm{LLT}(\lambda)\equiv\mu^\mrm{tr}(\lambda)\quad\text{and}\quad \mu^\mrm{T}(\lambda)\equiv \mu^\mrm{LTT}(\lambda) \, .
\eg
In particular, for $d=4$ explicit expressions for the spectral functions \cite{hyp1, Camporesi:1994} are 
\bg\label{eq:E21}
\spl{
\mu^\text{S}(\lambda) &= \frac{\pi}{16}\left(\lambda^2 +\frac{1}{4}\right)\lambda \tanh(\pi \lambda) \, ,\\
\mu^\text{T}(\lambda) &= \frac{\pi}{16}\left(\lambda^2 +\frac{9}{4}\right)\lambda \tanh(\pi \lambda) \, ,\\
\mu^\text{TT}(\lambda) &= \frac{\pi}{16}\left(\lambda^2 +\frac{25}{4}\right)\lambda \tanh(\pi \lambda) \,.
}
\eg
The relevant information on the spectrum of $-\Box_g$ on $\bbH^d(L)$ is gathered in Table~\ref{table:eigenvalues} below. 
\renewcommand*{\arraystretch}{1.5}
\begin{center}
	\begin{threeparttable}
		\begin{tabular}{|c|c|c|c|c|c|} \multicolumn{0}{c}{} \\
			\hline	{\textbf{Field}} & 
			{\textbf{Eigenfunction}} & 
			{\textbf{Eigenvalue}} & \textbf{Spectral function }
			\\ \hline\hline
			\underline{I = S}& $u_{\lambda l }^\text{S}(x)$ &$\mathscr{E}_\lambda ^\text{S}=\frac{\lambda^2+\rho^2}{L^2}$&$\mu^\text{S}(\lambda)$	\\ \hline
					\underline{I = V}&$\left(u_{\lambda l }^\text{T}\right)_\mu(x)$ &$\mathscr{E}_\lambda ^\text{T}=\frac{\lambda^2+\rho^2+1}{L^2}$&$\mu^\text{T}(\lambda)$	\\ 
			&$\left(u_{\lambda l }^\text{L}\right)_\mu(x)$ &$\mathscr{E}_\lambda ^\text{L}=\frac{\lambda^2+\rho(\rho+2)}{L^2}$&$\mu^\text{L}(\lambda)=\mu^\text{S}(\lambda)$	\\ \hline
			\underline{I = ST}			&$\left(u_{\lambda l }^\text{TT}\right)_{\mu\nu}(x)$ &$\mathscr{E}_\lambda ^\text{TT}=\frac{\lambda^2+\rho^2+2}{L^2}$&$\mu^\text{TT}(\lambda)$	\\
		&$\left(u_{\lambda l }^\text{LTT}\right)_{\mu\nu}(x)$ &$\mathscr{E}_\lambda ^\text{LTT}=\frac{\lambda^2+\rho(\rho+2)+3}{L^2}$&$\mu^\text{LTT}(\lambda)=\mu^\text{T}(\lambda)$\\ 
		 &$\left(u_{\lambda l }^\text{LLT}\right)_{\mu\nu}(x)$&$\mathscr{E}_\lambda^\text{LLT}=\frac{\lambda^2+\rho(\rho+4)+2}{L^2}$&$\mu^\text{LLT}(\lambda)=\mu^\text{S}(\lambda)$	\\
		 &$\left(u_{\lambda l }^\text{tr}\right)_{\mu\nu}(x)$&$\mathscr{E}_\lambda^\text{tr}=\frac{\lambda^2+\rho^2}{L^2}$&$\mu^\text{tr}(\lambda)=\mu^\text{S}(\lambda)$	\\ \hline
		\end{tabular}
		\caption{Decomposed eigenfunctions, associated eigenvalues and spectral functions, with $\lambda \in \bbR_{\geq 0}$ and $\rho=(d-1)/2$.}
		\label{table:eigenvalues}
	\end{threeparttable}
\end{center}

Next, consider the  operator trace-per-volume $\underline{\Tr}_{\mrm I}[\mcO]:=\Tr_{\mrm I}[\mcO]/\mrm{vol}[\mcM]$ on $\bbH^d(L)$ of a function $f(-\Box_g+c)$, where $c$ is an endomorphism of the  field species I, 
\begin{eqnarray}\label{trdef1}
	\underline{\Tr}_\text{I}[f(-\Box_g+c)]\is \frac{2^{d-2}\text{dim}(\rm I)}{\pi \Omega_{d-1}L^d}\int_0^\infty\!\! d\lb \,\mu^\text{I}(\lb) f(\mathscr{E}_\lambda^\text{I}+c)\,.
\end{eqnarray}
This formula may be justified by the following (formal though intuitive) computation, presented for the case of scalar fields for simplicity. The Laplacian $-\Box_g$ is essentially self-adjoint on $C_c^\infty(\bbH^d(L))$, and hence has a unique self-adjoint extension. It follows from the spectral theorem that 
\begin{eqnarray}\label{trdef2}
	\Tr_\text{S} [f(-\Box_g+c)]\is \int_0^\infty \!\!d\lb \sum_{l\in \mI(\text{S})}\int \!d^dx\,\sqrt{g}\,u_{\lb l}^\text{S}(x)^\ast f(-\Box_g+c)u_{\lb l}^\text{S}(x)
	\nonum
	\is \int_0^\infty \!\!d\lb \sum_{l\in \mI(\text{S})}\int \!d^dx\,\sqrt{g}\,u_{\lb l}^\text{S}(x)^\ast f(\mathscr{E}_\lambda^\text{S}+c)u_{\lb l}^\text{S}(x) 
	\nonum
	\is \int_0^\infty \!\!d\lb\,  f(\mathscr{E}_\lambda^\text{S}+c)\int \!d^dx\,\sqrt{g}\,\sum_{l\in \mI(\text{S})}u_{\lb l}^\text{S}(x)^\ast\cdot\, u_{\lb l}^\text{S}(x)
	\nonum
	\is \frac{2^{d-2}}{\pi \Omega_{d-1}}\frac{\mrm{vol}[\bbH^d(L)]}{L^d} \int_0^\infty \!\!d\lb\,\mu^\text{S}(\lb)  f(\mathscr{E}_\lambda^\text{S}+c)\,,  
\end{eqnarray}
where the final equality follows from the constancy (c.f. \eqref{E18a}) of $d^\text{S}_\lb=\sum_{l\in \mI(S)}u_{\lb l}^\text{S}(x)^\ast u_{\lb l}^\text{S}(x)$, together with the definition of the spectral function \eqref{eq:E16}. Then the trace-per-volume is given by dividing out the volume of $\bbH^d(L)$, yielding \eqref{trdef1} for $\text{I}=\text{S}$.\footnote{This argument can be formulated rigorously by considering the operator trace on the compact  manifold obtained by quotienting $\bbH^d(L)$ by a lattice of $SO(1,d)$, and subsequently analyzing the infinite volume limit (analogous to how a torus provides a volume regularization in flat spacetime).} Explicit expressions for the  relevant regularized traces are 
\begin{itemize}[leftmargin=2mm, rightmargin=-0mm]
\item[] {I = $\mrm S_N$}: 
\bg\label{eq:F6}
\spl{
\underline{\Tr}_{\text{S}_N}\left[f(-\Box_g+c)\right] &=  \frac{2^{d-2}}{\pi \Omega_{d-1}L^d}\int_0^N \!\! \text{d} \lambda\,\mu^\text{S}(\lambda) f\!\left(\mathscr{E}_\lambda^\text{S}+c\right)\,,
}
\eg
\item[] {I = $\mrm V_N$}:
\begin{align}\label{eq:F7}
	\underline{\Tr}_{\text{V}_N}[f(-\Box_g+c)] 
	 &=\frac{2^{d-2}}{\pi \Omega_{d-1}L^d}\int_0^N\!\! \text{d} \lambda\, \big\{(d-1)f\!\left(\mathscr{E}_\lambda^\text{T}+c\right) \mu^\text{T}(\lambda)+f\!\left(\mathscr{E}_\lambda^\text{L}+c\right) \mu^\text{S}(\lambda)\big\}\,,
\end{align}

\item[] {I = $\mrm{ST}_N$}:
\begin{align}\label{eq:F8}
	\underline{\Tr}_{\text{ST}_N}[f(-\Box_g+c)] &= \frac{2^{d-2}}{\pi \Omega_{d-1}L^d}\int_0^N\!\! \text{d} \lambda\, \bigg\{\frac{(d+1)(d-2)}{2}f(\mathscr{E}_\lambda^\text{TT}+c) \mu^\text{TT}(\lambda) \\[2mm]
	&+(d-1)f\!\left(\mathscr{E}_\lambda^\text{T}+c\right) \mu^\text{T}(\lambda) +\left[f\left(\mathscr{E}_\lambda^\text{LLT}+c\right)+f\!\left(\mathscr{E}_\lambda^\text{tr}+c\right)\right]\mu^\text{S}(\lambda)\bigg\} \, . \nonumber
\end{align}
\end{itemize}
Finally, the {\itshape degrees-of-freedom density}\footnote{More correctly, the quantity $\mff_{\rm I}(N)$ should be referred to as the ``degrees-of-freedom density per volume of the unit hyperboloid $\bbH^d(L=1)$''. However, for brevity we refer to it as the degrees-of-freedom density as above.}  $\mff_{\rm I}(N)$ defined as
\bg\label{eq:F10}
\mathfrak f_\mrm{I}(N):=L^d\underline{\Tr}_{\mrm I_N}[\mathds{1}_N]\, ,\quad \mrm I = \mrm S,\mrm V,\mrm{ST}\,,\quad N\in \bbR_{\geq 0}\,,
\eg
provides  a finite, $L$-independent count of `number of degrees of freedom' of field species I allowed by the $N$-cutoff regularization. 

%This diverges with $N$ as $\mathfrak f_\mrm{I}(N)\sim\int_0^N\D\lambda\,\mu^\mrm{I}(\lambda)$ in the limit $N\to\infty$.

%%%%%%%%%%%%%%%%%%%
%	New section 
%%%%%%%%%%%%%%%%%%%

\section{The free scalar field and quantum gravity on self-consistent hyperbolic backgrounds}
\label{sec:fields}

\subsection{The gravitational one-loop effective action induced from `matter' fields}

\noindent
\textbf{The scalar field.} Consider a free scalar field $A$ on $\bbH^d(L)$ which is subject to the action functional
\bg
\label{eq:D2}
S_\mrm{M}[A;g]=\foh\int\dd x\sg{x}\,A(x)\mK_x A(x) \, ,\quad \mK_x=-\Box_g^x+M^2+\xi R(x)\,,
\eg
where $M$ is the scalar field's mass and $\xi$ its non-minimal coupling to (classical) gravity. The gravity sector is governed by the Einstein-Hilbert action
\bg
\label{eq:D1}
S_\mrm{EH}[g]=\frac{1}{16\pi G}\int\dd x\sg{x}\,\lef(-R(x)+2\Lambda_\mrm{b}\ri) \, ,
\eg
with $G$ the (bare) Newton constant and $\Lambda_\mrm{b}$ the (bare) cosmological constant. Quantization of the scalar yields the following 
 gravitational one-loop effective action (which is exact for a free scalar)
\bg
\label{eq:D3}
\Gamma[g]:=\Gamma[A;g]\big|_{A=0}=S_\mrm{EH}[g]+\Gamma_\mrm{1L}[g] \, ,
\eg
where the one-loop term $\Gamma_\mrm{1L}[g]$ is obtained from the path integral
\bg
\label{eq:D4}
\e^{-\Gamma_\mrm{1L}[g]}=\intD\!\lef[g^{1/4}\what{A}\ri]\e^{-S_\mrm{M}[\what{A};g]}=\intD\what{B}\,\e^{-S_\mrm{M}[g^{-1/4}\what{B};g]} \, .
\eg
Here, $\mD[g^{1/4}\what{A}]=\mD\what{B}:=\prod_x\D\what{B}(x)$ is the diffeomorphism-invariant path integral measure and $B=g^{1/4} A$ the scalar field density whose action is
\bg\label{eq:D5}
S_\mrm{M}[g^{-1/4}B;g]=\foh\int\dd x\,B(x)\tilde\mK_x B(x) \, ,
\eg
with $\tilde\mK_x=g^{1/4}(x)\mK_x g^{-1/4}(x)=\mK_x $ since $-\Box_g^x$ arises from the Levi-Civita connection. 
The Gaussian path integral in \eqref{eq:D4} is evaluated in detail in Appendix~\ref{sec:PI}, and when regularized with an $N$-cutoff as in  \Gl{F6}, yields
\begin{eqnarray}\label{eq:D6}
	\Gamma_\mrm{1L}[g]_N \equiv \Gamma_{\text{1L}}(L)_N=\frac{1}{2}\Tr_{\text{S}_N}\log (\mathscr{K})= \frac{1}{2}\frac{2^{d-2}}{\pi \Omega_{d-1}} \frac{\text{vol}\left[\bbH^d(L)\right]}{L^d} \int_0^N\!\! \text{d}\lambda\, \mu^\text{S}(\lambda) \log \left(\mathscr{F}_\lambda^\text{S}(L)\right)\!,\,\,
\end{eqnarray}
with $\mathscr{F}_\lambda^\text{S}(L):=\mathscr{E}_\lambda^\text{S}(L)+M^2 + \xi R$.
\medskip

\textbf{Quantum Gravity.} Proceeding to  \ac{qg},  the full metric $g\mn$ is split into a background metric $\bar g\mn$ and a fluctuation part $h\mn$ according to the linear split,
\bg\label{eq:D7}
g\mn=\bar g\mn+h\mn \, .
\eg
The classical metrics $g\mn$ and $h\mn$ are to be understood as the expectation values (with respect to the path integral introduced below) of their quantum versions, $g\mn=\langle\what{g}\mn\rangle$ and $h\mn=\langle\what{h}\mn\rangle$, respectively. Further, the background metric $\bar g\mn$ stays purely classical, yet dynamical. Next, the two Faddeev-Popov ghost fields $\bar C\m=\langle\what{\bar C}\m\rangle$ and $C\M=\langle\what{C}\M\rangle$ are introduced in the standard way to give the full bare action
\bg\label{eq:D8}
S[h,\bar C,C;\bar g]=S_\mrm{EH}[\bar g+h]+S_\mrm{gf}[h;\bar g]+S_\mrm{gh}[h,\bar C,C;\bar g] \,.
\eg
Here,  $S_\mrm{EH}$ is the Einstein-Hilbert action \gl{D1}, while $S_\mrm{gf}$ and $S_\mrm{gh}$ are the gauge-fixing and ghost action respectively.\footnote{Since the full forms of $S_\mrm{gf}[h;\bar g]$ and $S_\mrm{gh}[h,\bar C,C;\bar g]$ are not explicitly required in this paper, we refer the interested reader to their precise forms given in \cite{Max2} (cf. Eqs. (4.2) and (18.2) in~\cite{MaxThesis}).}

Next, the  bare action \gl{D8} is expanded in $(h,\bar C,C)$ to yield the `matter' action,
\begin{align}\label{eq:D9}
S_\mrm{M}[h,\bar C,C;\bar g]&= S[h,\bar C,C;\bar g]-S_\mrm{EH}[\bar g]
\nonum 
&= S_\mrm{FG}[h;\bar g]+S_\mrm{Fgh}[\bar C,C;\bar g]+\mrm{(linear)}+O((h,\bar C,C)^3) \,.	
\end{align}
The terms linear in $(h,\bar C,C)$ do not contribute to the one-loop effective action by construction, thus these and terms of cubic order or higher in $(h,\bar C,C)$ can be omitted. This yields two terms, one quadratic in $h\mn$ and the other quadratic in $(\bar C,C)$, which will be interpreted as the actions for the free graviton ($\mrm{FG}$) and free ghost ($\mrm{Fgh}$) fields, respectively,
\begin{align}
	S_\mrm{FG}[h;\bar g]&=\foh\int\dd x\sgbo\,h\MN{\mK[\bar g]\mn}\RS h\rs \, , \label{eq:D10}\\
	S_\mrm{Fgh}[\bar C,C;\bar g]&=-\sqrt{2}\int\dd x\sgbo\,\bar C\m{\mM[\bar g,\bar g]\M}\n C\N \,.\label{eq:D11}
\end{align}
For any maximally symmetric spacetime\footnote{The results for the general case may be found in ~\cite{Max2}.} $(\mcM,g)$, in particular for $\bbH^d(L)$, the operators $\mK[\bar g]$ and $\mM[\bar g,\bar g]$ are given by\footnote{We have applied the harmonic gauge conditions, $\alpha=1$ and $\beta=1/2$, with $S_{\rm gf}$ as in \cite{Max2}.}
\bg\label{eq:D18}
\spl{
\mK[\bar g]=&\frac{1}{32\pi G}\lef(\mathds{1}_{\mrm{ST}^2}-\mathds{P}_\mrm{tr}\ri)\lef(-\bar D^2+2\Lambda_\mrm{b}+c_I\bar R\ri)\\
&-\frac{1}{32\pi G}\frac{d-2}{2}\mathds{P}_\mrm{tr}\lef(-\bar D^2+2\Lambda_\mrm{b}+c_\mrm{tr}\bar R\ri) \, ,
}
\eg
with $c_I=\frac{d(d-3)+4}{d(d-1)}$ and $c_\mrm{tr}=(d-4)/4$, as well as
\bg\label{eq:D19}
\mM[\bar g,\bar g]=\mathds{1}_\mrm{V}\lef(\bar D^2+\frac{1}{d}\bar R\ri) \, .
\eg
Here $\mathds{1}_{\mrm{ST}^2},\,\mathds{1}_\mrm{V}$ are the identity operators for the respective field species, and the projection operator $\mathds{P}_\mrm{tr}$ is defined in Appendix \ref{sec:deffields}.

In the subsequent section, we shall need the one-loop effective action for vanishing ghost arguments,
\bg\label{eq:D12}
\Gamma[h,0,0;\bar g]=S_\mrm{EH}[\bar g+h]+\Gamma_\mrm{1L}[\bar g+h]+O(\text{2 loops}) \, ,
\eg
with
\bg\label{eq:D13}
\Gamma_\mrm{1L}[g]=\Gamma_\mrm{FG}[g]+\Gamma_\mrm{Fgh}[g] \, .
\eg
The free graviton and free ghost fields' induced one-loop actions are obtained from the path integrals 
\bg\label{eq:D14}
\e^{-\Gamma_\mrm{FG}[\bar g]}=\intD\what{f}_{{\bigcdot}{\bigcdot}}\exp\lef(-S_\mrm{FG}\lef[\bar g^{-(d-4)/4d}\what{f}_{{\bigcdot}{\bigcdot}};\bar g_{{\bigcdot}{\bigcdot}}\ri]\ri) \, ,
\eg
and
\bg\label{eq:D15}
\e^{-\Gamma_\mrm{Fgh}[\bar g]}=\intD\what{B}^{{\bigcdot}}\mD\what{\bar B}_{\bigcdot}\exp\lef(-S_\mrm{FG}\lef[\bar g^{-(d-2)/4d}\what{\bar B}_{{\bigcdot}},\bar g^{-(d+2)/4d}\what{B}^{{\bigcdot}};\bar g_{{\bigcdot}{\bigcdot}}\ri]\ri)\,, 
\eg
where we have introduced the densitized fields $f\mn=\bar g^{(d-4)/4d}h\mn$, $\bar B\m=\bar g^{(d-2)/4d}\bar C\m$, and $B\M=\bar g^{(d+2)/4d} C\M$. Moreover, as for the scalar field,  the path integral measures are defined as $\mD(\cdots)=\prod_x(\cdots)(x)$, and are adjusted with factors of $\sgbo$ in order to make \gl{D14} and \gl{D15} \ac{brst}-invariant \cite{Fujikawa:2004cx}. The path integrals~\gl{D14} and \gl{D15} are merely Gaussian integrals, which when \ac{uv} regularized with the $N$-cutoff may be expressed in terms of  regularized traces (c.f. Eqs.~\gl{F7} and~\gl{F8}) as 
\bg\label{eq:D16}
\Gamma_\mrm{FG}[\bar g]_N=\foh\Tr_{\mrm{ST}_N}\log\mK[\bar g] \, ,
\eg
and
\bg\label{eq:D17}
\Gamma_\mrm{Fgh}[\bar g]_N=-\Tr_{\mrm{V}_N}\log\mM[\bar g,\bar g] \, .
\eg

%%%%%%%%%%%%%%%%%%%
%	New subsection 
%%%%%%%%%%%%%%%%%%%

\subsection{Equations of motion and the stress-energy tensor}
\label{ssec:set}

As discussed in the Introduction and in Section~\ref{sec:Ncutoff},   the principle of background independence \cite{Ashtekar:2014kba, Loll:2019rdj,Kie-Why, Casadio:2022ozp} is implemented in the form of a self-consistency condition at the level of the regularized system. Accordingly one must evaluate the backreaction of the quantized matter degrees of freedom on the (dynamical) background metric $g\mn\equiv\bar g\mn$, and thereby obtain the self-consistent solution $g_\munu ^\text{sc}$ from the resulting equations of motion. In case of the quantized free scalar field, the corresponding equations of motion read
\bg\label{eq:D20}
\frac{\delta}{\delta g\mn}\Gamma[g]_N\bigg|_{g\mn=g\mn^\mrm{sc}}=0 \, ,
\eg
where $\Gamma[g]_N$ is the $N$-cutoff regularized one-loop effective action~\gl{D3}. For quantized metric fluctuations, the corresponding equations of motion \cite{Isham-Prima, Giu-BI} are
\bg\label{eq:D22}
\frac{\delta}{\delta h\mn}\Gamma[h,0,0;\bar g]_N\bigg|_{h\mn=0,\bar g\mn=\bar g\mn^\mrm{sc}}=0 \, ,
\eg
where $\Gamma[h,0,0;\bar g]_N$ is the $N$-cutoff regularized version of the one-loop effective action for vanishing ghost arguments,\footnote{We select the $C=\bar{C}=0$ solution of the associated ghost tadpole equations $\langle \hat{C}\rangle =\langle \hat{\bar{C}}\rangle =0$ \cite{Max2}.} c.f. \Gl{D12}. 
Note that at one-loop order both equations of motion \gl{D20} and \gl{D22} may be expressed in the form\footnote{In going from \Gl{D22} to \Gl{D23}, we make use of the fact that at one-loop order, and vanishing fluctuation field, one may replace the functional derivative with respect to $h\mn$ with one with respect to $\bar g\mn$ \cite{Max2}.}
\bg\label{eq:D23}
\frac{\delta}{\delta g\mn}S_\mrm{EH}[g]\bigg|_{g\mn=g\mn^\mrm{sc}}+\frac{\delta}{\delta g\mn}\Gamma_\mrm{1L}[g]_N\bigg|_{g\mn=g\mn^\mrm{sc}}=0 \, ,
\eg
where the regularized one-loop corrections are given by \Gl{D6} in case of a quantized free scalar field, and \Gl{D13} (together with Eqs.~\gl{D16} and \gl{D17}) in case of quantized metric fluctuations. We stress that the regularization scheme that enters the equations of motion \gl{D23} must be strictly background independent, with no (geometric) background structure  fixed `by hand'. The  $N$-cutoff regularization precisely allows   for the  self-consistent metric $g\mn$ to be determined in this way. We further remark that while the $N$-cutoff regularized  one-loop effective actions for both the scalar and gravity sectors contain a volume divergence due to $\mrm{vol}[\bbH^d(L)]$, the equations of motion \gl{D25}, \gl{D26} below are rendered fully finite by the $N$-cutoff, c.f. \gl{D29b}, \gl{D29c}.

Defining the effective (regularized) \ac{set} following  Euclidean conventions as
\bg\label{eq:D24}
T\MN[g]_N:=-\frac{2}{\sgo}\frac{\delta}{\delta g\mn}\Gamma_\mrm{1L}[g]_N \, ,
\eg
 the equations of motion~\gl{D23} take the familiar form of the Einstein equation,
\bg\label{eq:D25}
R\mn-\foh g\mn R+\Lambda_\mrm{b}g\mn=8\pi G T\mn[g]_N \, .
\eg
On maximally symmetric spacetimes, such as $\bbH^d(L)$, the \ac{lhs} of \Gl{D25} is
\begin{eqnarray}
	\bigg[\frac{1}{d}\bigg(-\frac{d}{2}+1\bigg)R+\Lambda_\mrm{b}\bigg]g\mn\,,
\end{eqnarray}
with $R\equiv\text{constant}$. Since the sub-bases defining the $N$-cutoffs (c.f. \gl{F5}) have been chosen such that they carry a representation of the background's symmetry group, it follows that $T_\munu[g]_N$ is proportional to $g_\munu$ with an $N$-dependent proportionality factor. Therefore, it is sufficient to consider the traced version of \gl{D25},
\bg\label{eq:D26}
\lef(-\frac{d}{2}+1\ri)R+d\Lambda_\mrm{b}=8\pi G T\M\m[g]_N \,,
\eg
with $T\M\m[g]_N$ an $N$-dependent constant on $\bbH^d(L)$. 

To proceed we note the simple identity\footnote{It is readily checked that this is valid beyond in the non-maximally symmetric case as well.}
\begin{eqnarray}\label{trform}
	\int\! \D^d x \sqrt{g}g^\munu T_\munu[g]_N\is \frac{\D}{\D\alpha} {\Gamma_{1\rm L}[e^{-2\alpha} g]}_N\bigg|_{\alpha = 0}\,,
\end{eqnarray}
which follows from the definitions of the \ac{set} and functional derivative. The variation of the operator trace(s)  (cf. Eq.~(17.9) in~\cite{MaxThesis}) appearing in $\Gamma_{1\rm L}[g]_N$ for scalars \gl{D6}   yields 
\begin{eqnarray}\label{eq:D29b}
	T^\mu _\mu [g]_N^\text{scal} &=& \underline{\Tr}_{\mrm S_N}\lef[\frac{-\Box_g+\xi R}{-\Box_g+M^2+\xi R}\ri]\nonumber\\	
&=&\frac{2^{d-2}}{\pi \Omega_{d-1}}\frac{1}{L^d}\int_0^N\!d\lb \,\mu^\text{S}(\lb)\bigg[1-\frac{M^2}{\mathscr{E}_\lb^\text{S}(\lb)+M^2+\xi R}\bigg] \,,
\end{eqnarray}
while the result for  metric fluctuations follows from  Eqs.~\gl{D13}, \gl{D16} and \gl{D17},
\begin{eqnarray}\label{eq:D29c}
	T^\mu _\mu [g]_N^\text{grav}=\underline{\Tr}_{\mrm{ST}_N}\lef[\mK[g]^{-1}\lef(\mK[g]\big|_{\Lambda_\mrm{b}=0}\ri)\ri]-2\underline{\Tr}_{\mrm{V}_N}\lef[\mathds{1}_\mrm{V}\ri] \, ,
\end{eqnarray}
with $\underline{\Tr}_{\mrm I_N}[\mcO]$ the (regularized) trace-per-volume of the operator $\mcO$, c.f. Eq. \eqref{trdef1}.

It is clear from Eqs. \gl{D29b}, \gl{D29c} that the self-consistency condition for both scalars and metric fluctuations is fully regularized by the $N$-cutoff, in contrast to the associated one-loop effective actions. Although it is instructive to focus on the effective action due to the central role it plays in continuum approaches to quantum gravity, e.g. the asymptotic safety scenario \cite{Reuter:2019byg}, we remark that the (fully regularized) self-consistency relations may instead be obtained directly by a path integral average of its classical counterpart. Indeed, for the scalar field\footnote{We emphasize that the \ac{rhs} of \Gl{D30} does not correspond to a quantized version of the typically employed \ac{set}, which classically is given by $-\frac{2}{\sgo}\frac{\delta}{\delta g\mn}S_\mrm{M}[A; g]$, due to the extra $g$-dependence in the first argument of $S_\mrm{M}$ in \Gl{D30}. Classically, both quantities agree on-shell (though not off-shell), while their quantum versions differ by a divergent term. It has been argued that \gl{D30} constitutes the more natural choice for a quantum \ac{set}~\cite{Max2}. The discussion for the case of metric fluctuations follows analogously.}
\bg\label{eq:D30}
T\MN[g]_N=-\frac{1}{\sgo}\frac{\delta}{\delta g\mn}\lef\langle S_\mrm{M}\lef[g^{-1/4}\what{B}; g\ri]\ri\rangle_N \, ,
\eg
where $S_\mrm{M}$ is given by \gl{D5} and the expectation value is taken with respect to path integral of \Gl{D4}, regularized with the $N$-cutoff. This yields the traced \ac{set} in terms of 
 the densitized eigenfunctions $v_{\lambda l}^\text{S}=g^{1/4}u_{\lambda l}^\text{S}$,
\begin{align}\label{eq:F12}
	T\M\m[g]_N^{\rm scal}&=\frac{1}{\sgo}\int_0^N\!\! \D\lambda\,\sum_l\frac{1}{\mF_\lambda^\mrm{S}}\Bigg\{\left(1-\frac{d}{2}\right)D\m v_{\lambda l}^{\text{S}\,\ast}D\M v_{\lambda l}^\text{S}
-v_{\lambda l}^{\text{S}\,\ast} D^2v_{\lambda l}\Bigg.
\nonum 
&\Bigg.+\xi R|v_{\lambda l}^\text{S}|^2-2\xi(1+d) v_{\lambda l}^{\text{S}\,\ast} D^2v_{\lambda l}^\text{S}-2d\xi D\m v_{\lambda l}^{\text{S}\,\ast} D\M v_{\lambda l}^\text{S} 
 \Bigg\} \, ,
\end{align}
with  $\mF_\lambda = \mE_\lambda + M^2 +\xi R$, which is rendered entirely finite by the $N$-cutoff. The equivalence of the two expressions \gl{D29b} and \gl{F12} for the traced \ac{set} follows from the discussion in Section \ref{ssec:2_3} and results of Appendix \ref{sec:maths}.

\subsection{Self-consistent hyperbolic $N$-geometries}
\label{subsec:Ngeometries}

Having developed the $N$-cutoff regularization scheme for the scalar and metric fluctuations on  $\bbH^d(L)$, culminating in the regularized \ac{set}s \gl{D29b} and  \gl{D29c} expressed in terms of the  operator traces \gl{F6}-\gl{F8}, we now evaluate the backreaction of the regularized field's modes via the contracted Einstein equation~\gl{D26}, i.e.,
\bg\label{eq:G1}
\left(-\frac{d}{2}+1\right)R(L) + d \Lambda_\text{b} = 8\pi G\,T\m\M(L)_N \, ,
\eg
with $T\m\M(L)_N\equiv T\m\M[g=L^2\gamma]_N$ and $R(L)=-{d(d-1)}/{L^2}$. Solutions to \Gl{G1} are the self-consistent hyperbolic background geometries. In the presence of the cutoff $N$, we will refer to these as `\ita{$N$-geometries}'.

%%%%%%%%%%%%%%%%%%%
%	New subsubsection 
%%%%%%%%%%%%%%%%%%%

\subsubsection{The free scalar field}

It is convenient to reexpress the traced \ac{set} \gl{D29b} of a free scalar field
%
% The traced \ac{set} of a free scalar field, \Gl{D29b}, regularized by an $N$-type cutoff according to \Gl{F6}, i.e.,
%\bg\label{eq:G2}
%T^\mu_\mu (L)^\text{scal}_N =  \underline{\Tr}_{\text{S}_N}\left[\frac{-\Box_g + \xi R(L)}{-\Box_g +M^2 +\xi R(L)}\right]\, .
%\eg
%We can rewrite \Gl{G2} 
in terms of the degrees-of-freedom density, c.f.  \Gl{F10},
\begin{align}\label{eq:G3}
	T^\mu_\mu (L)^\text{scal}_N &= \frac{1}{L^d}\mathfrak{f}_\text{S}(N)+\Delta T^\mu_\mu(L;M)^\text{scal}_N \, ,
\end{align}
with
\begin{subequations}
\begin{align}\label{eq:G6}
\mathfrak{f}_\text{S}(N) &= \frac{2^{d-2}}{\pi \Omega_{d-1}}\int_0^N \text{d}\lambda\, \mu^\text{S}(\lambda)\,, 
\\[2mm]
\label{eq:G4}
	\Delta T^\mu_\mu(L;M)^\text{scal}_N &= - \underline{\Tr}_{\text{S}_N}\left[\frac{M^2}{-\Box_g +M^2 +\xi R(L)}\right]\,.
\end{align}
\end{subequations}
We note that, in particular, for a massless scalar field one has $\Delta T^\mu_\mu(L;M)^\text{scal}_N = 0$. 

Next, inserting \Gl{G3} into the equations of motion \gl{G1} yields
\bg\label{eq:G5}
\left(-\frac{d}{2}+1\right)R(L) + d \Lambda_\text{b} = 8\pi G\left\{\frac{1}{L^d} \mathfrak{f}_\text{S}(N)+\Delta T^\mu_\mu(L;M)^\text{scal}_N \right\} \, ,
\eg
Specializing to $d=4$ spacetime dimensions, $\mathfrak{f}_\text{S}(N)$  behaves as
\bg\label{eq:G6}
\mathfrak{f}_\text{S}(N) =\frac{1}{32\pi^2}N^4 \left(1+O\lef(1/N\ri)\right)\,, \quad N\to \infty\,.
\eg
 It is intriguing to point out that $\mathfrak f_\mrm{S}(N)$ diverges  quartically in this case, with the infamous logarithmic divergences  only subleading on $\bbH^d(L)$ (and, for comparison, not present at all in the case $\mcM=\bbS^d(L)$~\cite{Max1}).

\medskip
 
\begin{figure}[h]
	\centering
	\includegraphics[scale=0.4]{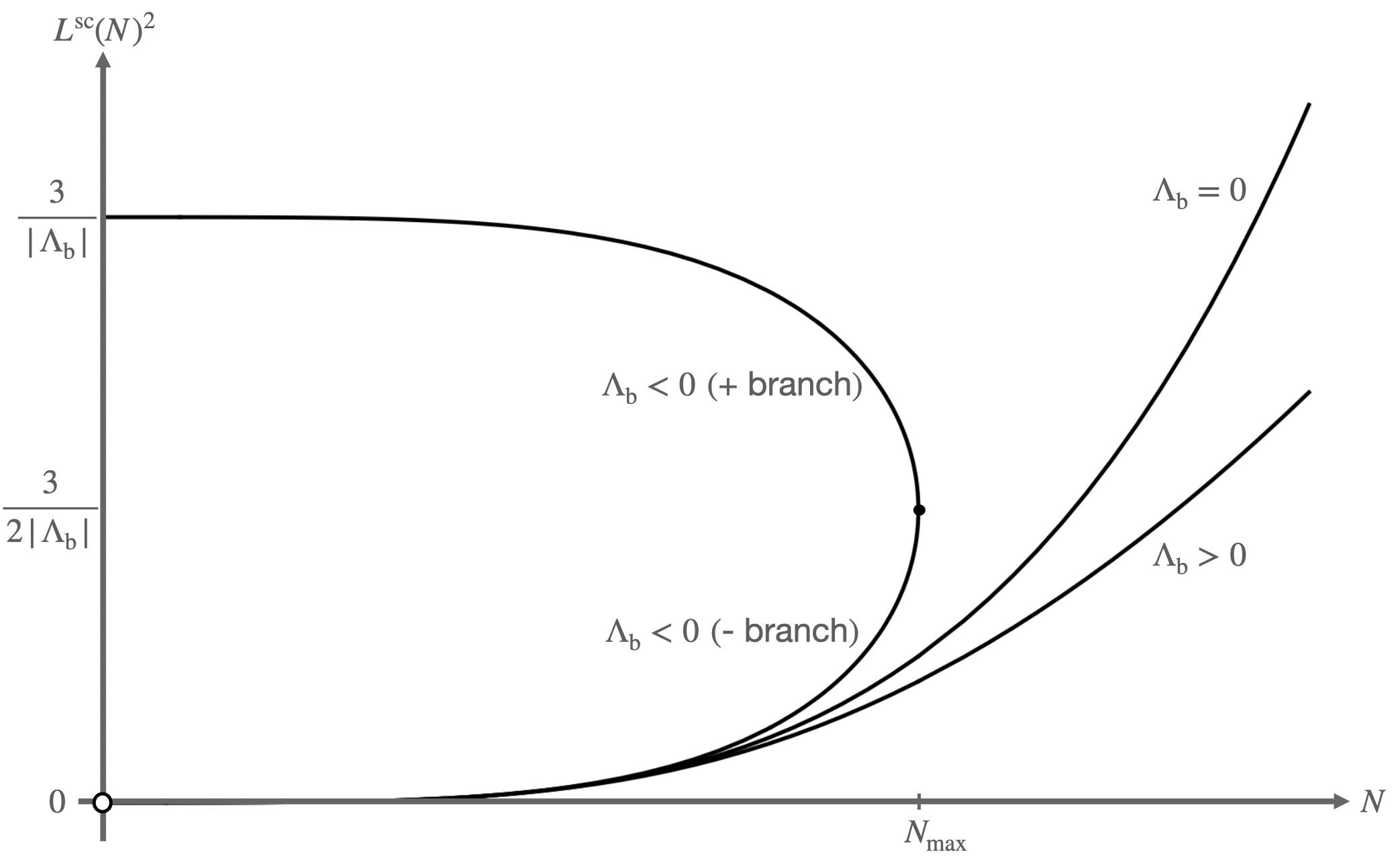}
	\caption{Self-consistent hyperbolic $N$-geometries for $\Lambda_\mrm{b}=0$, $\Lambda_\mrm{b}>0$, and $\Lambda_\mrm{b}<0$.}
	\label{fig1}
\end{figure}

\textbf{The massless scalar field in $d=4$ spacetime dimensions.} For the case $M=0$ and $d=4$, the equations of motion~\gl{G1} with the \ac{set}~\gl{G3} reduce to a simple equation for $L^2$,
\bg
\label{eq:G7}
\Lambda_\mrm{b}L^4+3L^2-2\pi G\,\mathfrak{f}_\mrm{S}(N)=0 \, ,
\eg
whose general solution for the self-consistent radius squared $L^\mrm{sc}(N)^2$ is given by
\begin{subequations}
	\begin{align}\label{eq:G8a} 
	L^\mrm{sc}(N)^2&=\frac{2\pi G}{3}\mathfrak f_\mrm{S}(N)\,,\quad \Lambda_\mrm{b}= 0\,,
	\\[3mm]\label{eq:G8b}
	L^\mrm{sc}(N)^2 &=\frac{3}{2\Lambda_\mrm{b}}\lef\{-1\pm\sqrt{1+\frac{8\pi G\Lambda_\mrm{b}}{9}\mathfrak f_\mrm{S}(N)}\ri\}\,,\quad \Lambda_\mrm{b}\neq 0\,.
\end{align}
\end{subequations}
These solutions have distinct features depending on the sign of the bare cosmological constant, and are depicted in Figure \ref{fig1} and discussed below.

\textbf{(i) The case $\Lambda_\mrm{b}=0$.} It follows from   \Gl{G8a} that there is a self-consistent hyperbolic $N$-geometry for each $N>0$,  and the large $N$ asymptotic behavior  \gl{G6} yields $L^\mrm{sc}(N)^2\sim N^4$ as $N\to \infty$. Thus, one observes the striking result that {\itshape including increasingly many field modes tends to drive the self-consistent curvature to zero}, leading to a \ita{perfectly flat geometry} in the \ac{qft} limit $N\to\infty$. This is precisely the opposite behavior of what is usually perceived as the ``cosmological constant problem'', where adding modes leads to a diverging curvature. On the other hand, since $\mff_\mrm{S}(N)$ vanishes as $N\to 0^+$, the self-consistent radius vanishes in this limit, yielding a curvature singularity. Thus, the existence of the family of self-consistent $N$-geometries is a purely quantum effect, lacking a well-defined limit in the absence of quantum modes, i.e., the $N\to 0^+$ limit.

%{}with no classical $(N\to 0^+)$ limit.

%For a vanishing bare cosmological constant, \Gl{G8} converges (via the `$+$'-branch) to the simple expression
%\bg
%\label{eq:G9}
%L^\mrm{sc}(N)^2=\frac{2\pi G}{3}\mathfrak f_\mrm{S}(N) \, .
%\eg
%Consequently, the self-consistent radius scales as $L^\mrm{sc}(N)\propto N^2$ and we observe striking phenomenon that \ita{when adding further quantized field modes, the resulting curvature reduces}, leading to a \ita{perfectly flat geometry} in the \ac{qft} limit $N\to\infty$. This is precisely the opposite behavior of what is usually perceived in the ``cosmological constant problem'', where adding modes increases curvature. Additionally, in the solution~\gl{G9} does not have a classical limit since $N\to 0$ yields to a curvature singularity. Thus, this solution is purely quantum. Also, among the possible maximally-symmetric background $N$-geometries, there seem to exist only negative curvature solutions for the case $M=0$ and $\Lambda_\mrm{b}=0$, because for this case no spherical solution exists~\cite{Max1}.

\textbf{(ii) The case $\Lambda_\mrm{b}>0$.} When the bare cosmological constant is non-zero, the self-consistency equation \gl{G7} is quadratic in $L^2$, and hence there may be two $(\pm)$ branches for the solution $L^\mrm{sc}(N)^2$ in \gl{G8b}. For strictly positive $\Lambda_\mrm{b}$, however, one obtains self-consistent $N$-geometries with positive  $L^\mrm{sc}(N)^2$  only for the `+' branch in \Gl{G8b}. Indeed, the existence of \ita{hyperbolic} background $N$-geometries for all $N>0$ with  a \ita{positive} bare cosmological constant is remarkable, and is a purely quantum effect without a well-defined  limit in the absence of quantum modes since $L^\mrm{sc}(N)^2\to 0$ as $N\to 0^+$. Moreover, it follows from the large $N$ asymptotics \gl{G6} that the squared-radius {\itshape diverges as} $L^\mrm{sc}(N)^2\sim N^2$ as $N\to \infty$, and hence the cosmological constant problem stemming from summing up vacuum energies is entirely absent.

%As in the above $\Lambda_\mrm{b}=0$, one finds self-consistent solutions  $L^\mrm{sc}(N)^2$ for all $N>0$, and it follows from the large $N$ asymptotics \gl{G6} that the squared-radius {\itshape diverges as} $L^\mrm{sc}(N)^2\sim N^2$ as $N\to \infty$.

%For a positive bare cosmological constant, the \ac{rhs} of \Gl{G8} is strictly positive for $N>0$ when choosing the `+'-branch. For $N\to 0$ we again encounter a curvature singularity, such that this solution is purely quantum. The existence of \ita{hyperbolic} background $N$-geometries for a \ita{positive} bare cosmological constant is remarkable for itself. Then, the self-consistent radius scales asymptotically as $L^\mrm{sc}(N)\sim N$ and qualitatively we find the very same notable behavior as before: $L^\mrm{sc}(N)\to\infty$ for $N\to\infty$. Thus, also in this case, the cosmological constant problem stemming from summing up vacuum energies is fully absent. For the comparison with spherical backgrounds, in this case the qualitatively same solution has been found, which however connects to the classical solution $L=3/\Lambda_\mrm{b}$ for $N\to 0$~\cite{Max1}.

\textbf{(iii) The case $\Lambda_\mrm{b}<0$.} In the case of negative bare cosmological constant, it is natural to expect that self-consistent hyperbolic $N$-geometries exist for $N\geq 0$. To proceed, it is convenient to reexpress \gl{G8b} as
\bg
\label{eq:G10}
L^\mrm{sc}(N)^2=\frac{3}{2|\Lambda_\mrm{b}|}\lef\{1\mp\sqrt{1-\frac{8\pi G|\Lambda_\mrm{b}|}{9}\mathfrak f_\mrm{S}(N)}\ri\} \, ,
\eg
with $\Lambda_\mrm{b}=-|\Lambda_\mrm{b}|$. Thus one has solutions for both `$\pm$' branches with, however, the surprising result that  hyperbolic space with negative bare cosmological constant can self-consistently support quantum fluctuations only up to a {\itshape maximum value} $N=N_\mrm{max}$ defined by 
\begin{align}\label{eq:G10a}
	\mff_\text{S}(N_\mrm{max}):=\frac{9}{8\pi G |\Lambda_\text{b}|}\,,
\end{align}
i.e., the \ac{qft} limit $N\to \infty$  does not exist. Further, when the combination $G|\Lambda_{\rm b}|$ is small, the large $N$ asymptotics \gl{G6} may be employed to  solve  \Gl{G10a} yielding the maximum value
\begin{align}
\label{eq:G11}
N_\mrm{max}\approx\sqrt[4]{\frac{36\pi}{G|\Lambda_\mrm{b}|}} \,.
\end{align}
Finally, the two branches of solutions in \Gl{G10} may be characterized as follows. For the `$-$' branch there is a  curvature singularity at $N=0$ (i.e. the existence of this family of solutions is a purely quantum effect), with  $L^\mrm{sc}(N)^2$ increasing  with $N$ up to the maximal value $L^\mrm{sc}(N_\mrm{max})$. On the other hand, for the `$+$' branch one begins at  the classical hyperbolic space solution $L^\mrm{sc}(N=0)^2=-3/\Lambda_\mrm{b}$, with the radius-squared decreasing with $N$ up to a minimum value $L^\mrm{sc}(N_\mrm{max})^2$. 
%The two branches of solutions in \Gl{G10} then can be characterized as follows. Choosing the `$-$'-branch, we again find a curvature singularity for $N=0$, i.e., there are only quantum solutions here. For $N>0$ the solution for $L^\mrm{sc}(N)$ then increases linearly up to $L^\mrm{sc}(N_\mrm{crit})$. On the other hand, choosing the `$+$'-branch, we recover the classical hyperbolic solution $L=-3/\Lambda_\mrm{b}$ in the limit $N\to 0$. From this classical value, $L^\mrm{sc}(N)$ then decreases linearly down to $L^\mrm{sc}(N_\mrm{crit})$, mirroring the solution of the other branch (except for $N=0$).

\medskip

\textbf{The massive scalar field in $d=4$ spacetime dimensions.} In the case of a non-vanishing mass $M$ of the scalar field, the effect of the $\Delta T^\mu_\mu(L;M)^\text{scal}_N $ term on the 
 \ac{rhs} of \Gl{G5} must be accounted for. Explicitly from \Gl{G4} one has
\bg\label{eq:G12}
\Delta T^\mu_\mu(L;M)^\text{scal}_N=-\frac{4}{\pi\Omega_3 L^4}\int_0^N\!\!\D\lambda\,\mu^\mrm{S}(\lambda)\frac{M^2}{\mE_\lambda^\mrm{S}(L)+M^2+\xi R(L)} \, .
\eg
Since $\mE_\lambda^\mrm{S}\sim\lambda^2$ and $\mu^\mrm{S}(\lambda)\sim\lambda^3$ for $\lambda$ large, $\Delta T^\mu_\mu(L;M)^\text{scal}_N$ will only diverge quadratically in $N$ and is thus subleading compared to the quartic divergence of $\mathfrak f_\mrm{S}(N)$. Hence, in the {\itshape large $N$ regime}, the conclusions drawn in the $M=0$ case remain valid. In particular, for $\Lambda_\mrm{b}\geq 0$ there is a family of self-consistent hyperbolic $N$-geometries with vanishing curvature in the limit $N\to \infty$, i.e. yielding a flat geometry in the \ac{qft} limit.

%%%%%%%%%%%%%%%%%%%
%	New subsubsection 
%%%%%%%%%%%%%%%%%%%

\subsubsection{Metric fluctuations (quantum gravity)}

 Next, consider the self-consistent  $N$-geometries arising from metric fluctuations  on hyperbolic space (i.e., from their induced \ac{ol} effective gravitational action). The  traced \ac{set} entering the \ac{rhs} of the contracted Einstein equation \gl{G1} is given by \Gl{D29c}, expressed in terms of the $N$-cutoff regularized traces~\gl{F7}-\gl{F8}\footnote{Technically, each sector I of the decomposition of ST and V could be assigned its own cutoff value $N_\mrm{I}$. Therewith, different minimal lengths $L/N_\mrm{I}$ could be resolved in each sector. However, here we choose the case $N_\mrm{I}\equiv N$.}, explicitly
\begin{align}\label{eq:G13}
	T^\mu_\mu(L)^\text{grav}_N&=\underline{\Tr}_{\mrm{ST}_N}\lef[\frac{\mK(L)|_{\Lambda_\mrm{b}=0}}{\mK(L)}\ri]-2\underline{\Tr}_{\mrm V_N}\lef[\mathds{1}_\mrm{V}\ri]\\[2mm] 
&=\underline{\Tr}_{\mrm{ST}_N}\lef[\lef(\mathds{1}_{\mrm{ST}^2}-\mathds{P}_\mrm{tr}\ri)\frac{-\Box_g+c_I R(L)}{-\Box_g+2\Lambda_\mrm{b}+c_I R(L)}+\mathds{P}_\mrm{tr}\frac{-\Box_g+c_\mrm{tr} R(L)}{-\Box_g+2\Lambda_\mrm{b}+c_\mrm{tr} R(L)}\ri]\nonum 
&\ \ \ -2\underline{\Tr}_{\mrm V_N}\lef[\mathds{1}_\mrm{V}\ri] \, ,\nonumber
\end{align}
with the second equality following from \Gl{D18}. As in the scalar case, it is convenient  to express this in terms of the degrees-of-freedom densities,
\bg\label{eq:G14}
T^\mu_\mu(L)^\text{grav}_N=\frac{1}{L^d}\lef\{\mff_\mrm{FG}(N)-\mff_\mrm{Fgh}(N)\ri\}+\Delta T^\mu_\mu(L;\Lambda_\mrm{b})^\text{grav}_N \,,
\eg
where the degrees-of-freedom densities for the free graviton (FG) and free ghost fields (Fgh) are, respectively,
\begin{align}\label{eq:G15}
\mff_\mrm{FG}(N)=\mff_{\mrm{ST}}(N)=L^d\underline{\Tr}_{\mrm{ST}_N}[\mathds{1}_{\mrm{ST}}]\,,\quad \mff_\mrm{Fgh}(N)=2\mff_{\mrm{V}}(N)=2L^d\underline{\Tr}_{\mrm{V}_N}[\mathds{1}_{\mrm{V}}] \, .
\end{align}
It is intriguing to observe that evaluating Eq.~\gl{G15} with the hyperbolic $N$-traces~\gl{F6}-\gl{F8} yields to leading order in $N$,
\bg\label{eq:G16}
\mff_\mrm{FG}(N)-\mff_\mrm{Fgh}(N)=\lef[\foh(d+1)(d-2)-(d-1)\ri]\mff_\mrm{S}(N)+O\lef(N^{d-1}\ri) \, ,
\eg
i.e., precisely the $d(d-3)/2$ polarization states of the gravition field. The very same result had already been found in the spherical case~\cite{Max2}. 

Next, the last term on the \ac{rhs} of \Gl{G14} is given by
\begin{align}
	\label{eq:G17}
\Delta T^\mu_\mu(L;\Lambda_\mrm{b})^\text{grav}_N=-\underline{\Tr}_{\mrm{ST}_N}&\bigg[\lef(\mathds{1}_{\mrm{ST}^2}-\mathds{P}_\mrm{tr}\ri)\frac{2\Lambda_\mrm{b}}{-\Box_g+2\Lambda_\mrm{b}+c_I R(L)}
\nonum 
&+\mathds{P}_\mrm{tr}\frac{2\Lambda_\mrm{b}}{-\Box_g+2\Lambda_\mrm{b}+c_\mrm{tr} R(L)}\bigg]\, .
\end{align}
For the very same reasons as outlined in the scalar case, namely that $\mE_\lambda^\mrm{I}\sim\lambda^2$ and $\mu^\mrm{I}(\lambda)\sim\lambda^{d-1}$ for large $\lambda$, it is evident from the $N$-trace fomulae \gl{F6}-\gl{F8}, that $\Delta T^\mu_\mu(L;\Lambda_\mrm{b})^\text{grav}_N$ will only give subleading contributions in $N$. 

We consider here only the large $N$-behavior, in which case the  traced \ac{set}  is given by
\bg\label{eq:G18}
T^\mu_\mu(L)^\text{grav}_N=\frac{1}{L^d} \frac{d(d-3)}{2}\mff_\mrm{S}(N)+O\lef(N^{d-1}\ri) \,.
\eg
Therefore, in the large $N$ regime, the self-consistent  hyperbolic $N$-geometries of the graviton fluctuations coincide with those of the massless scalar at leading order in $N$. In particular, this implies that the $N\to \infty$ behavior of the $N$-geometries for the graviton field may be obtained from the corresponding results \gl{G7}-\gl{G10} for the free scalar under the  substitution $\mathfrak f_\mrm{S}(N)\to 2\mathfrak f_\mrm{S}(N)$  in $d=4$ dimensions. For $\Lambda_\text{b}<0$, there are self-consistent hyperbolic $N$-geometries only up to some $N_\text{max}<\infty$. On the other hand, for $\Lambda_\text{b}\geq 0$, there are solutions $L^\text{sc}(N)^2$ for all $N>0$ with the property that the associated (negative) curvature tends to zero in the \ac{qft} limit $N\to \infty$, i.e., the cosmological constant problem is absent.

%Accordingly, for $\Lambda_\text{b}\geq 0$ the resulting (negative) curvature tends to zero in the \ac{qft} limit $N\to \infty$, and for $\Lambda_\text{b}<0$ modes up to some $N_\text{max}<\infty$ can be self-consistently quantized on the hyperbolic space background.

% This implies that \ita{the hyperbolic $N$-geometries for a scalar field with vanishing mass and for the graviton field at leading order in $N$ are qualitatively the  same}. Indeed, it is clear  that the $N\to \infty$ behavior of the $N$-geometries for the graviton field may be obtained from the corresponding results \gl{G7}-\gl{G10} for the free scalar under the  substitution $\mathfrak f_\mrm{S}(N)\to 2\mathfrak f_\mrm{S}(N)$  in $d=4$ dimensions.

%%%%%%%%%%%%%%%%%%%
%	New section 
%%%%%%%%%%%%%%%%%%%

\section{Conclusion}
\label{sec:conclusion}

The self-consistent determination of any background structures appearing in a quantum theory is an essential conceptual feature in the presence of non-trivial gravitational interactions, whether in semi-classical or full quantum gravity. The $N$-cutoffs are a self-consistent quantization scheme in which background independence is imposed at the level of the regularized precursor of the QFT, and has been developed here for scalar and metric fluctuations on the non-compact hyperbolic space. For both field species there are self-consistently determined hyperbolic $N$-geometries with the remarkable feature that, without any fine-tuning of parameters, the inclusion of increasingly many quantum modes does not lead to the usual cosmological constant problem of a curvature singularity. Instead they show the striking behavior that the hyperbolic backgrounds become flatter as $N$ is increased, ultimately leading to a perfectly flat background in the \ac{qft} (i.e., $N\to \infty$) limit. The details have already been discussed in Section \ref{subsec:Ngeometries}, and will not be repeated here. Instead, we comment on some extensions and future directions.

First, it is of interest to explore  the  self-consistent quantization via $N$-cutoffs formalism  on Lorentzian signature manifolds, for which a natural setting is to analyze fields on the maximally symmetric de Sitter and  anti-de Sitter spacetimes. Another important extension is to study the consequences of the formalism on manifolds with less than maximal symmetry, with (Euclidean signature) product manifolds such as $\bbS^1(L_1)\times\bbS^3(L_2)$ being a potential setting (wherein  a Wick rotation from Euclidean to  Lorentzian signature has already  been studied~\cite{Grosse:2011es}). 
%A natural extension of the $N$-cutoff formalism would be to fields on maximally symmetric ({\itshape Lorentzian} signature) de Sitter  and anti-de Sitter spacetimes.   A related  setting would be to analyze self-consistent product spacetimes, such as $\bbS^1(L_1)\times\bbS^3(L_2)$, where a Wick rotation from Euclidean to  Lorentzian signature has  been studied~\cite{Grosse:2011es}.
% As a second step, it would be instructive to look into interacting fields, e.g., a scalar field with a potential. 

 Another important direction is to study the effect of interactions  by relating $N$-cutoff regularization to a renormalization scheme,
 %Another important direction is to relate the $N$-cutoff regularization to a renormalization scheme
 possibly by implementing a renormalization group flow equation. Although the $N$-cutoff is by construction scale-free, this approach does generate a self-consistent scale, namely the  hyperboloid's radius $L^\mrm{sc}(N)$, which in turn defines a self-consistent momentum scale $\mcP^{\rm sc}(N)=(N^2+\rho^2)^{1/2}/L^\mrm{sc}(N)$ for e.g. scalars. For each $N$ this places an upper bound on the accessible momenta, and it is interesting to note that $\lim_{N\to \infty}\mcP^{\rm sc}(N)<\infty$, namely even in the \ac{qft} limit there remains an upper limit on the momenta of the fields. Relatedly, analogous results have been observed in attempts to connect the renormalization group scale employed in the \ac{frg} framework to the spectral flow of the quantized field modes~\cite{Pagani:2019vfm, Ferrero:2022hor}. 
However, the relation of this upper bound on the accessible momenta to a genuine adjustable renormalization scale remains to be clarified. 
%Unfortunately, it is unclear how this relates to a genuine adjustable renormalization scale. 
Further, since flow equations such as the Polchinski equation \cite{Polchinski:1983gv} are constructed from the requirement of scale invariance of the path integral in presence of a dimensionful \ac{uv} cutoff, it is an open question if the renormalization group \`a la Wilson can be implemented solely from the $N$-cutoff point of view. Nonetheless, an $N$-cutoff renormalization scheme would provide a useful tool for analyzing further physical implications of the $N$-cutoffs, e.g. by making contact with traditional results from \ac{qft} in curved spacetime, such as the \ac{set}'s trace anomaly \cite{Parker:1978gh,Fulling:1989nb, Birrell:1982ix, Brown:1976wc, Morris:2018zgy}.

{\bfseries Acknowledgements}

We thank Martin Reuter, Max Niedermaier, Frank Saueressig, Roberto Percacci, Dario Benedetti, Roberto Camporesi, and  Rainer Verch for interesting discussions. The authors are also grateful to  Martin Reuter, Max Niedermaier,  Frank Saueressig, and Roberto Percacci 
for helpful comments on the manuscript.   
The authors thank the Institut Henri Poincar\'e (UAR 839 CNRS-Sorbonne Universit\'{e}) for  hospitality at the workshop ``Quantum gravity, random geometry and holography'', during which part of this work was completed. The author acknowledge support of the Institut Henri Poincar\'{e} (UAR 839 CNRS-Sorbonne Universit\'{e}), and LabEx CARMIN (ANR-10-LABX-59-01).
MB gratefully acknowledges financial support by the Deutsche Forschungsgemeinschaft (DFG, German Research Foundation) -- project number 493330310.

%%%%%%%%%%%%%%%%%%%
%%%%%%%%%%%%%%%%%%%
%	Appendix
\appendix
%%%%%%%%%%%%%%%%%%%
%%%%%%%%%%%%%%%%%%%

%%%%%%%%%%%%%%%%%%%
%	New section 
%%%%%%%%%%%%%%%%%%%

\section{Matter fields on Riemannian background spacetimes}
\label{sec:deffields}

\noindent
In this paper, scalar ($\mrm S$), vector ($\mrm V$) and symmetric rank-2 tensor fields ($\mrm{ST}$) appear, and are often collectively denoted by the label $\mrm I$. We note that in general, tensor fields are covariant fields by default (i.e., they only possess lower indices in local coordinate/abstract index notation). The corresponding Hilbert spaces of square-integrable fields\footnote{The local trivialization of the $\mrm V$ and $\mrm{ST}^2$ vector bundles have fibres $V_x=\{x\}\times\bbC^d$ and $\mrm{ST}_x=\{x\}\times\bbC^{d(d+1)/2}$. The associated Hilbert spaces can thus be locally represented in terms of the scalar Hilbert space as $L^2_{\mrm S}(\mcM,g)=L^2((\mcM,g);\bbC)$, $L^2_{\mrm V}(\mcM,g)=L^2_{\mrm S}(\mcM,g)\otimes\bbC^d$, and $L^2_{\mrm{ST}}(\mcM,g)=L^2_{\mrm S}(\mcM,g)\otimes\bbC^{d(d+1)/2}$.}, $L_{\mrm I}^2(\mcM,g)$ with $\mrm I=\mrm S, \mrm V, \mrm{ST}$, may be regarded as the completion of the space of smooth, compactly supported sections of the associated vector bundles with respect to the canonical inner products
\bg
\label{eq:B1}
\spl{
	\langle\tilde{\sigma}\,{,}\,\sigma\rangle_\text{S}&\!:=\!\int\!d^dx \sqrt{g(x)}\,\tilde{\sigma}(x)^\ast \sigma(x)\,,
	\\[2mm]
	\langle\tilde{\xi}\,{,}\,\xi\rangle_\text{V}&\!:=\! \int\!d^dx \sqrt{g(x)}\,g^\munu(x)\tilde{\xi}_\mu(x)^\ast \xi_\nu (x)\,,
	\\[2mm]
	\langle\tilde{\tau}\,{,}\,\tau\rangle_\text{ST}&\!:=\! \int\!d^dx \sqrt{g(x)} \,I[g]^{\mu \nu \rho \sigma}(x)\tilde{\tau}_{\mu \nu}(x)^\ast \tau_{\rho\sigma}(x)\,,\quad I[g]\MNRS={\textstyle \frac{1}{2}}(g\MR g\NS+g\MS g\NR)\,.}
\eg
Unlike scalar fields, vector and tensor fields may be decomposed further. Vector fields may be decomposed into transverse ($\mrm I=\mrm T$) and longitudinal ($\mrm I=\mrm L$) pieces, while symmetric rank-2 tensor fields may be split into the four parts of the York decomposition\footnote{The vector field and York decomposition apply only to tensors which are defined on Einstein manifolds.} ($\mrm I=\mrm{TT}, \mrm I =\mrm{LTT},\mrm I=\mrm{LLT},\mrm I=\mrm{tr}$) \cite{York}.  Furthermore, it is not hard to see that these decompositions yield mutually orthogonal subspaces under the inner products given in Eq.~\gl{B1}.\\

\noindent
The following decompositions are valid for vector fields and symmetric rank-2 tensor fields on any Einstein manifold.
%%%%%%%%%%%%%%%%%%%%%%%%%%%%%%%%%%%%%%%%%%%%
\paragraph{Decomposition of $\mrm V$.}
The Hilbert space $L^2_\mrm{V}(\mcM,g)$ can be decomposed as
\bg
\label{eq:C1}
L^2_\mrm{V}(\mcM,g)=L^2_\mrm{T}(\mcM,g)\oplus L^2_\mrm{L}(\mcM,g)\ ,\ \xi\m=\xi\m^{\mrm T}+\xi\m^{\mrm L} \, ,
\eg
with the transverse part $\xi\m^{\mrm T}$ of the vector field $\xi\m$ defined by $D\M\xi\m^{\mrm T}=0$. The associated scalar products $\langle\,\bigcdot\,{,}\,\bigcdot\,\rangle_\mrm{T}$ and $\langle\,\bigcdot\,{,}\,\bigcdot\,\rangle_\mrm{L}$ arise from the corresponding restriction of $\langle\,\bigcdot\,{,}\,\bigcdot\,\rangle_\mrm{V}$ to $\mrm T$ and $\mrm L$. Locally, we may identify
\bg
\label{eq:C2}
L^2_\mrm{T}(\mcM,g)\cong L^2_\mrm{S}(\mcM,g)\otimes\bbC^{d-1}\quad\text{and}\quad L^2_\mrm{L}(\mcM,g)\cong L^2_\mrm{S}(\mcM,g) \, ,
\eg
where the latter isomorphism is given by the mapping
\bg
\label{eq:C3}
P^{\mrm L,\mrm S} : L^2_\mrm{S}(\mcM,g)\to L^2_\mrm{L}(\mcM,g)\ ,\ \sigma\mapsto (P^{\mrm L,\mrm S})\m\sigma:= D\m\frac{1}{\sqrt{-D^2}}\sigma=\xi\m^{\mrm L} \, .
\eg
It is easy to verify that the mapping \gl{C3} defines an {\itshape isometry} of the spaces $L^2_\mrm{S}(\mcM,g)$ and $L^2_\mrm{L}(\mcM,g)$, with  $\langle\sigma_1\,{,}\,\sigma_2\rangle_{\mrm S}=\langle\xi_1^\mrm{L}\,{,}\,\xi_2^\mrm{L}\rangle_\mrm{L}$ for ${\xi_i^\mrm{L}}_\mu=D\m\frac{1}{\sqrt{-D^2}}\sigma_i$, $i=1,2$.\footnote{
We remark that often the isomorphism $L^2_\text{S}(\mcM,g)\simeq L^2_\text{L}(\mcM,g)$ is defined via the local mapping $\xi\m^\mrm{L}=D\m\sigma$, in contrast to the non-local definition employed here. It is important to note that while the mapping \eqref{eq:C3} has the  advantage of turning the relationship between $\mrm S$ and $\mrm L$ into an isometry,  this transformation is not legitimate for physical degrees of freedom. However, $\sigma$ itself is a gauge degree of freedom \cite{Percacci:2017fkn}. 
}
This property is employed in this paper to construct an orthonormal basis for $\mrm L$ from a given one for  $\mrm S$  in the case of $\mcM$ maximally symmetric.

Finally, we denote the projectors onto $\mrm T$ and $\mrm L$ by $\bbP_\mrm{T}$ and $\bbP_\mrm{L}$, which have matrix elements
\bg
\label{eq:C5}
\langle{x,\mu}|\bbP_\mrm{I}|{y,\nu}\rangle=({{P_\mrm{I}}\m}\N)^{\mrm{diff},x}\delta(x-y)/\sg{y}\ ,\ \mrm I=\mrm T,\mrm L \, .
\eg
The explicit form of the projectors ${{P_\mrm{I}}\m}\N$ as differential operators is not relevant here and can be found e.g. in \cite{Benedetti:2010nr, Groh:2011dw}.

%%%%%%%%%%%%%%%%%%%%%%%
\paragraph{Decomposition of $\mrm{ST}$.}
First, we decompose the Hilbert space $L^2_{\mrm{ST}}(\mcM,g)$ as
\bg
\label{eq:C6}
L^2_{\mrm{ST}}(\mcM,g)=L^2_{\mrm{TT}}(\mcM,g)\oplus L^2_{\mrm{LT}}(\mcM,g)\oplus L^2_{\mrm{tr}}(\mcM,g)\ ,\ \tau\mn=\tau\mn^\mrm{TT}+\tau\mn^\mrm{LT}+\tau\mn^\mrm{tr} \, .
\eg
Here, $\mrm{TT}$ is the space of transverse-traceless tensor fields, i.e., $D\M\tau\mn^\mrm{TT}=0$ and $g\MN\tau\mn^\mrm{TT}=0$. Locally, we identify
\bg
\label{eq:C7}
L^2_{\mrm{TT}}(\mcM,g)\cong L^2_{\mrm{S}}(\mcM,g)\otimes\bbC^{(d+1)(d-2)/2} \, .
\eg
Next, $\mrm{LT}$ is the space of longitudinal-traceless tensor fields. Since locally we have,
\bg
\label{eq:C8}
L^2_{\mrm{LT}}(\mcM,g)\cong L^2_{\mrm{S}}(\mcM,g)\otimes\bbC^d=L^2_{\mrm{V}}(\mcM,g) \, ,
\eg
we can further decompose $L^2_{\mrm{LT}}(\mcM,g)$ by means of the vector field decomposition \gl{C1},
\bg
\label{eq:C9}
L^2_{\mrm{LT}}(\mcM,g)=L^2_{\mrm{LTT}}(\mcM,g)\oplus L^2_{\mrm{LLT}}(\mcM,g)\ ,\ \tau\mn^\mrm{LT}=\tau\mn^\mrm{LTT}+\tau\mn^\mrm{LLT} \, .
\eg
$\mrm{LTT}$ is the space of longitudinal-transverse traceless tensor fields, locally isomorphic to
\bg
\label{eq:C10}
L^2_{\mrm{LTT}}(\mcM,g)\cong L^2_{\mrm{S}}(\mcM,g)\otimes\bbC^{d-1}\cong L^2_{\mrm{T}}(\mcM,g) \, .
\eg
Thus we can identify $\mrm{LTT}$ and $\mrm{T}$ via the mapping
\begin{align}\label{eq:C11}
	&(P^{\mrm{LTT},\mrm{T}})\n :  \,L^2_{\mrm{T}}(\mcM,g)\to L^2_{\mrm{LTT}}(\mcM,g)\ ,\\[2mm] 
	&\xi\m^\mrm{T}\mapsto (P^{\mrm{LTT},\mrm{T}})\n\xi\m^\mrm{T}=\frac{1}{\sqrt{2}}\lef[ D\m\frac{1}{\sqrt{-D^2-\text{Ric}}}\xi\n^\mrm{T}+D\n\frac{1}{\sqrt{-D^2-\text{Ric}}}\xi\m^\mrm{T}\ri]=\tau\mn^\mrm{LTT}\,,\nonumber
\end{align}
with $(\text{Ric}\,\xi)_\mu = R_\mu\, ^\nu \xi_\nu$.
Again, it is straightforward to see that definition with \gl{C11}, the spaces $\mrm T$ and $\mrm{LTT}$ become isometric.

Lastly, we can locally identify
\bg
\label{eq:C12}
L^2_{\mrm{LLT}}(\mcM,g)\cong L^2_{\mrm{S}}(\mcM,g)\quad\text{and}\quad L^2_{\mrm{tr}}(\mcM,g)\cong L^2_{\mrm{S}}(\mcM,g) \, ,
\eg
and the mappings that make the Hilbert spaces of $\mrm{LLT}$ and $\mrm{tr}$ isometric to that of $\mrm S$, respectively, are given by\footnote{
As in case for the decomposition of $\mrm V$, here all nonlocal field redefinitions are legitimate since only gauge degrees of freedom are concerned.
}
\begin{align}\label{eq:C13}
	&(P^{\mrm{LLT},\mrm{S}})\mn : \ L^2_{\mrm{S}}(\mcM,g)\to L^2_{\mrm{LLT}}(\mcM,g)\ ,\\[2mm]  
&\sigma\mapsto (P^{\mrm{LLT},\mrm{S}})\mn\sigma=\sqrt{\frac{d}{d-1}}\lef(D\m D\n-\frac{1}{d}g\mn D^2\ri)\frac{1}{\sqrt{(-D^2)^2+\frac{d}{d-1} D_\mu R^\munu D_\nu }}\sigma=\tau\mn^\mrm{LLT} \, ,\nonumber 
\end{align}
%
%\bg
%\label{eq:C13}
%\spl{
%(P^{\mrm{LLT},\mrm{S}})\mn : &\ L^2_{\mrm{S}}(\mcM,g)\to L^2_{\mrm{LLT}}(\mcM,g)\ ,\\ 
%&\sigma\mapsto (P^{\mrm{LLT},\mrm{S}})\mn\sigma=\sqrt{\frac{d}{d-1}}\lef(D\m D\n-\frac{1}{d}g\mn D^2\ri)\frac{1}{\sqrt{-D^2}}\frac{1}{\sqrt{-D^2-R/(d-1)}}\sigma=\tau\mn^\mrm{LLT} \, ,
%}
%\eg
and
\bg
\label{eq:C14}
(P^{\mrm{tr},\mrm{S}})\mn : L^2_{\mrm{S}}(\mcM,g)\to L^2_{\mrm{tr}}(\mcM,g)\ ,\
\sigma\mapsto (P^{\mrm{tr},\mrm{S}})\mn\sigma=\frac{1}{\sqrt{d}}g\mn=\tau\mn^\mrm{tr} \, .
\eg
We denote the projectors on the subspaces of the York decomposition by $\bbP_\mrm{I}$, $\mrm I=\mrm{TT},\mrm{LTT},\mrm{LLT},\mrm{tr}$, with the matrix elements
\bg
\label{eq:C15}
\langle{x,\mu,\nu}|\bbP_\mrm{I}|{y,\rho,\sigma}\rangle=({{\bbP_\mrm{I}}\mn}\RS)^{\mrm{diff},x}\delta(x-y)/\sg{y} \, .
\eg
For the specific form of the differential operators ${{\bbP_\mrm{I}}\mn}\RS$ see e.g.~\cite{Benedetti:2010nr, Groh:2011dw}.

%%%%%%%%%%%%%%%%%%%
%	New section 
%%%%%%%%%%%%%%%%%%%

\section{Addition theorems on hyperbolic space}
\label{sec:maths}

\begin{theorem}[Hyperbolic harmonics addition theorem]
	\ \label{thadd}	
Consider the hyperboloid $\bbH^d(L)$ and associated (scalar) eigenfunctions $u_{\lb l}$ satisfying $-\Box_{\bbH^d(L)}u_{\lb l}=\mathscr{E}_\lb u_{\lb l}$, with $\lb\in \bbR_{\geq 0}$ and $l$ a discrete multi-index taking values in a countable set of infinite cardinality. 
	Given $\lb\geq 0$, define the function $d_\lb:\bbH^d(L)\to [0,\infty]$, 
\begin{eqnarray}\label{ad1}
	d_\lb(x):=\sum_{l}u_{\lb l}(x)u_{\lb l}(x)^\ast \,,\quad \forall\,x\in \bbH^d(L)\,.
\end{eqnarray}
%where we use the standard coordinatization $x=(y,\Omega)$ of $\bbH^d(L)$, c.f. \Gl{E2}.
Then $d_\lb$ is a finite constant function, 
\begin{eqnarray}
	d_\lb(x)\equiv \frac{2^{d-2}}{\pi \Omega_{d-1}L^d}\mu(\lambda)\,, \quad\forall x\in \bbH^d(L)\,,
\end{eqnarray}
with $\mu(\lb)$ the spectral function for scalars. We will write (by slight abuse of notation) $d_\lb(x)\equiv d_\lb$.
\end{theorem}
Theorem \ref{thadd} is the analogue of the addition theorem for spherical harmonics, Theorem \ref{thadd1} below. In proving Theorem  \ref{thadd}, the subtleties that arise  when dealing with the {\itshape continuous spectrum} of  $-\Box_{\bbH^d(L)}$ on the {\itshape non-compact} hyperbolic space are discussed. Since we focus only on the scalar case, for notational brevity the field species label $\text{I}=\text{S}$ will be omitted below, i.e.,  we will write $u_{\lb l}\equiv u_{\lb l}^\text{S}$, $\mathscr{E}_\lb \equiv \mathscr{E}_\lb^\text{S}$, $d_\lb \equiv d_\lb^\text{S}$, and $\mu_\lb \equiv \mu_\lb^\text{S}$.

\begin{theorem}[Addition theorem for spherical harmonics]
	\ \label{thadd1} 
Consider the $D$-dimensional unit sphere $\bbS^D$, and the associated (scalar) spherical harmonics $Y_{\ell m}$ satisfying $-\Box_{\bbS^D}Y_{\ell m}=\ell(\ell+D-1)Y_{\ell m}$, where $\ell \in \bbN_0$ and $m$ is a multi-index labeling the  finitely many orthogonal eigenfunctions in the $\ell^\text{th}$ eigenspace. Then for all $\Omega \in \bbS^D$
\begin{eqnarray}\label{ad2a}
	\sum_{m}|Y_{\ell m}(\Omega)|^2\is \frac{G_\ell}{{\rm vol}(\bbS^D) }\,,\quad {\rm vol}(\bbS^D) = \frac{2 \pi^{(D+1)/2}}{\Gamma((D+1)/2)}\,,
\end{eqnarray}
and 
\begin{eqnarray}\label{ad2b}
	G_\ell := \frac{(2 \ell+D-1)(\ell+D-2)!}{\ell!(D-1)!}\,
\end{eqnarray}
is the dimension of the $\ell^\text{th}$ eigenspace.	
\end{theorem}

{\itshape Proof of the addition Theorem for (scalar) spherical harmonics (Theorem \ref{thadd1}):}
The Laplacian $-\Box_{\bbS^D}$ is essentially self-adjoint on $C_c^\infty(\bbS^D)$, and hence has a unique self-adjoint extension in $L^2(\bbS^D)$. Its inverse is a compact operator, thereby having a fully discrete spectrum. The spectral theorem then entails that the set of spherical harmonics $\{Y_{\ell m}\}$ forms a complete orthonormal basis of $L^2(\bbS^D)$. Given two arbitrary points $\Omega,\,\Omega' \in \bbS^D$, the transitive action of $SO(D)$ implies that there is a group element $\sigma$ such that $\Omega'=\sigma \vartriangleright \Omega$.\footnote{Here `$\vartriangleright$' denotes the usual left-action of the symmetry group on the manifold.} Then, clearly $Y_{\ell m}(\sigma \vartriangleright \Omega)=Y_{\ell m}\circ \sigma (\Omega)$. Since $\sigma$ is an  isometry of the sphere, the set $\{Y_{\ell m}\circ \sigma \}$ is also a complete orthonormal basis of $L^2(\bbS^D)$, and hence the two bases may be expressed in terms of one another,
%Then clearly the set $\{Y_{\ell m}\circ \sigma \}$ also constitute a complete orthonormal basis of $L^2(\bbS^D)$. The two bases are related by a unitary mapping
\begin{eqnarray}\label{ad12a}
	Y_{\ell m}\circ \sigma =\ssum{\ell' m'}{} \alpha_{\ell m,\ell'm'}(\sigma)Y_{\ell' m'}\,,\quad \sum_{\ell''m''}\alpha_{\ell''m'',\ell m}(\sigma)\alpha_{\ell''m'',\ell 'm'}(\sigma)^\ast =\delta_{\ell \ell'}\delta_{mm'}\,,
\end{eqnarray}
with the second relation following from the unitarity of the basis change map $L^2(\bbS^D)\to L^2(\bbS^D)$. Next, to conclude that only $\ell'=\ell$ terms appear in the above expansion, we note that the action of $-\Box_{\bbS^D}$ commutes with $\sigma \in SO(D)$, which implies that $\alpha_{\ell m,\ell'm'}(\sigma)=\delta_{\ell \ell'}C_{\ell mm'}(\sigma)$. Inserted into the expansion \eqref{ad12a}, this eliminates the sum over $\ell'\in \bbN_0$, leaving only the sum over $m$ that consists of finitely many terms,
\begin{eqnarray}
	Y_{\ell m}\circ \sigma =\ssum{ m'}{} C_{\ell mm'}(\sigma)Y_{\ell m'}\,.
\end{eqnarray}
Further, the unitarity relation for the coefficients $\alpha_{\ell m,\ell'm'}(\sigma)$ in \eqref{ad12} implies that the coefficients $C_{\ell mm'}(\sigma)$ satisfy $\sum_{m''}C_{\ell m''m}(\sigma)C_{\ell m''m'}(\sigma)^\ast =\delta_{mm'}$ for all $\ell \in \bbN_0$. Hence we have
%Thus $Y_{\ell m}\circ \sigma$ may be expressed in terms of $Y_{\ell m}$ through a {\itshape finite} sum, whose coefficients satisfy $\sum_{m''}C_{\ell m''m}(\sigma)C_{\ell m''m'}(\sigma)^\ast =\delta_{mm'}$. 
%
%Hence
%\red{Add: $\delta_{\ell\ell'}$ eliminates sum over $l$ $\Rightarrow$ only sum over $m$ remains, which is over a finite set. And $\sum_{m''}C_{\ell m''m}(\sigma)C_{\ell m''m'}(\sigma)^\ast =\delta_{mm'}$ holds or all $\ell$/lhs is independent of $\ell$}
\begin{eqnarray}\label{ad13}
	\sum_{m}|Y_{\ell m}(\sigma \vartriangleright \Omega)|^2\is \sum_{m}|Y_{\ell m}\circ\sigma(\Omega)|^2
	= \sum_{m}\sum_{m',m''}Y_{\ell m'}(\Omega)Y_{\ell m''}(\Omega)^\ast C_{\ell m m'}(\sigma)C_{\ell m m''}(\sigma)^\ast 
	\nonum
	\is \sum_{m',m''}Y_{\ell m'}(\Omega)Y_{\ell m''}(\Omega)^\ast \sum_{m}C_{\ell m m'}(\sigma)C_{\ell m m''}(\sigma)^\ast
	= \sum_{m}|Y_{\ell m}(\Omega)|^2\,,\qquad
\end{eqnarray}
where the order of finite sums may readily be exchanged. Therefore we conclude that $\sum_{m}|Y_{\ell m}(\Omega)|^2$ is a constant on the sphere.
The final result \eqref{ad2a} is obtained by 
integrating this constant over the sphere and using the orthonormality of the spherical harmonics.
\qed

{\bfseries Remark.} In order to transition from the sphere to hyperbolic space, it is helpful to observe that this argument consists of two key components:
%This argument may be distilled into two components: 
(i) The transitivity of the group action entails that for any two points on the manifold, there is a group element mapping between them. (ii) For any group element $\sigma$, $Y_{\ell m}\circ \sigma$ is itself an element of $L^2(\bbS^D)$, and hence by the spectral theorem may be expanded in the $Y_{\ell m}$ with coefficients satisfying a unitarity relation \eqref{ad12}. Further, the summations appearing in the calculation \eqref{ad13} involve only finitely many terms, and therefore require no additional statements on convergence.
It is clear that (i) generalizes readily to hyperbolic space, where indeed the action of $SO(1,d)$ can link any two points $x,\,x'\in \bbH^d(L)$. Unfortunately, the argument of (ii) does not generalize so readily. The culprit of course is that the  Laplacian has a purely continuous spectrum on $\bbH^d(L)$. The eigenfunctions $u_{\lb l}$ do not belong to $L^2(\bbH^d(L))$, and hence the spectral theorem cannot be directly applied to conclude that $u_{\lb l}\circ \sigma $ (for $\sigma\in SO(1,d)$) may be expressed in terms of the $u_{\lb l}$ themselves.\footnote{There is no tension between the fact that eigenfunctions $u_{\lb l}\notin L^2(\bbH^d(L))$, and the statement that they constitute a complete orthonormal basis for $L^2(\bbH^d(L))$. {\itshape Completeness} and {\itshape orthonormality} are to be understood in terms of the spectral theory of $-\Box_{\bbH^d(L)}$ whereby $L^2(\bbH^d(L))$ is unitarily equivalent to a Hilbert space in which the Laplacian acts by multiplication. This unitary mapping is determined by the $u_{\lb l}$ via 
\begin{eqnarray*}
	f(x)\mapsto f^\wedge(\lb ,l):=\int_{\bbH^d(L)}\!\!\!d^dx \sqrt{g(x)}u_{\lb l}(x)^\ast f(x)\,,\quad f\in C_c^\infty(\bbH^d(L))\subseteq L^2(\bbH^d(L))\,.
\end{eqnarray*}
} Moreover, the precise sense of the convergence of such an expression is unclear (e.g. it cannot be relative to the $L^2$-norm, since neither $u_{\lb l}$ nor $u_{\lb l}\circ \sigma $ are square integrable). While such questions can be addressed through  Nuclear Spectral Theory, we find it instructive to present the following more elementary analysis. 

\medskip
 Recall that $-D_\mu D^\mu u_{\lb l}=\mathscr{E}_\lb(L) u_{\lb l}$ with 
\begin{eqnarray}\label{ad3}
	\mathscr{E}_\lb(L)= (\lb^2 + \rho^2)/L^2\,,\quad \rho:=(d-1)/2\,.
\end{eqnarray}
 In standard coordinates $(y,\Omega)$ on $\bbH^d(L)$, with $y\geq 0$ and $\Omega\in \bbS^{d-1}$,
\begin{eqnarray}\label{ad4}
	u_{\lb \ell m}(y,\Omega)\is N_{\lb \ell}(L)\,q_{\lb \ell}(y)\,Y_{\ell m}(\Omega)\,, \quad  \lb\in \bbR_{\geq 0}\,,\quad \ell\in \bbN_0\,,
\end{eqnarray}
where $m$ distinguishes different eigenfunctions with eigenvalues  labeled by $\lb$ and $\ell$.\footnote{We have  decomposed the degeneracy multi-index $l$ into $(\ell, m)$. As is the case for the  sphere, for each $\ell$,  $m$ takes values in a finite set.} Here
$Y_{\ell m}$ are the complete orthonormal scalar  spherical harmonics on $\bbS^{d-1}$, 
\begin{eqnarray}\label{ad5}
	-\Box_{\bbS^{d-1}}Y_{\ell m}=\ell(\ell+d-2)Y_{\ell m}\,.
\end{eqnarray}
Next,  $q_{\lb \ell}$ is conveniently expressed in terms of a 
hypergeometric function,
\begin{eqnarray}\label{ad6}
	q_{\lb \ell}(y):=(i\,\sinh y)^\ell F\Big(i\lb+\rho+\ell,\,-i\lb+\rho+\ell;\,\ell+ \frac{d}{2};-\sinh^2(y/2)\Big)\,,
\end{eqnarray}
and the normalization factor is 
\begin{eqnarray}\label{ad7}
	N_{\lb \ell}(L)=\frac{1}{L^{d/2}}\bigg|\frac{2^{ \ell -(d-2)/2}\Gamma( \ell +d/2)\Gamma(i\lb)}{\Gamma(i\lb +\rho+ \ell )}\bigg|\inv\,.
\end{eqnarray}

In order to prove Theorem \ref{thadd} we prepare the following lemmata.
\begin{lemma}
	\ \label{Adlm1}
	{\itshape For every $x\in \bbH^d(L)$, and every $\lb\geq 0:\,d_\lb(x)=\sum_{l}|u_{\lb l}(x)|^2<\infty$.}
\end{lemma}
{\bfseries Remark.} This is weaker than the statement of Theorem \ref{thadd} in that it only asserts that $d_\lb$ is a finite function on  $\bbH^d(L)$, with its constancy remaining unproven.
\begin{lemma}
	 \label{Adlm2}
{\itshape	
	Given $\sigma\in SO(1,d)$ and an eigenfunction $u_{\lb l}$, consider the composition $u_{\lb l}\circ \sigma$. Then there is a pointwise expansion
\begin{eqnarray}\label{ad12}
		(u_{\lb l}\circ \sigma) (x)\is \ssum{j}{}C_{\lb lj}(\sigma) u_{\lb j}(x)\,,
\end{eqnarray}
which further converges uniformly for the compact sets $[y_0,y_1]\times \bbS^d\subseteq \bbH^d(L)$ with $y_0>0$. Moreover, the coefficients $C_{\lb  l j}(\sigma)$ are unitary matrices (in the $j,\,l$ indices), satisfying
\begin{eqnarray}
	\ssum{k}{}C_{\lb k j}(\sigma)C_{\lb k l}(\sigma)^\ast =\delta_{jl}\,.
\end{eqnarray}}
\end{lemma}

{\itshape Proof of the addition theorem for (scalar) hyperbolic harmonics (Theorem \ref{thadd}):}
Fix an arbitrary $x\in \bbH^d(L)$, and define $x_0\in \bbH^d(L)$ as the point on the hyperboloid at with coordinate $y=0$. Then there is $\sigma\in SO(1,d)$ such that $x=\sigma\vartriangleright x_0$. Thus by Lemma \ref{Adlm2}
\begin{eqnarray}\label{ad20}
	d_\lb (x)\is \sum_{l}|u_{\lb l}\circ \sigma(x_0)|^2=\sum_{l}\sum_{j,\,j'}C_{\lb lj}(\sigma)C_{\lb lj'}(\sigma)^\ast u_{\lb j}(x_0)u_{\lb j'}(x_0)^\ast 
	\nonum
	\is \sum_{j,\,j'}u_{\lb j}(x_0)u_{\lb j'}(x_0)^\ast\sum_{l}C_{\lb lj}(\sigma)C_{\lb lj'}(\sigma)^\ast
	\nonum
	\is \sum_{j,\,j'}u_{\lb j}(x_0)u_{\lb j'}(x_0)^\ast \delta_{jj'}\,,
\end{eqnarray}
i.e., $d_\lb(x)=d_\lb (x_0)$. Note that since $u_{\lb j}(x_0)\neq 0$ for only finitely many $j$ (as may be seen by inspection of the explicit expression \eqref{ad6} at $x_0$, see also \eqref{ad4}), there is no difficulty in exchanging the orders of summation in arriving at the penultimate equality. Hence $d_\lb$ is a constant function on the hyperboloid, with its relation to the spectral function following by comparing to the defining relation \eqref{eq:E16}.
\qed 

\medskip

We note the useful corollary

\begin{corollary}
	\ \label{coradd}
 For all $x\in \bbH^d(L)$ and $\lb\geq 0$, $\sum_{l}D_\mu u_{\lb l}(x)D^\mu u_{\lb l}(x)^\ast=-\mathscr{E}_\lb d_\lb \,.$
\end{corollary}

{\itshape Proof of Corollary \ref{coradd}:}
Fixing $x\in \bbH^d(L)$ arbitrary, with $x_0$ the point with coordinate $y=0$, by a calculation analogous to \eqref{ad20} we arrive at
\begin{eqnarray}
	\sum_{l}D_\mu u_{\lb l}(x)D^\mu u_{\lb l}(x)^\ast \is \sum_{l}D_\mu u_{\lb l}(x_0)D^\mu u_{\lb l}(x_0)^\ast
	\nonum
	\is \half  \sum_l D_\mu D^\mu |u_{\lb l}(x_0)|^2-\sum_l  u_{\lb l}(x_0)^\ast D^\mu D_\mu u_{\lb l}(x_0)\,.
%	\nonum
%	\is \half D_\mu D^\mu\Big(\sum_l |u_{\lb l}(x_0)|^2\Big)-\mathscr{E}_\lb \sum_l |u_{\lb l}(x_0)|^2
%	\nonum
%	\is -\mathscr{E}_\lb  d_\lb \,.
\end{eqnarray}
Since $D_\mu D^\mu |u_{\lb l}(x_0)|^2\neq 0$ for only finitely many $l$, the derivative may be taken outside the summation, yielding $D_\mu D^\mu d_\lb =0$. This leaves $\sum_{l}D_\mu u_{\lb l}(x)D^\mu u_{\lb l}(x)^\ast =-\mathscr{E}_\lb  d_\lb$, completing the proof.
\qed

Finally, we prove Lemmas \ref{Adlm1} and \ref{Adlm2}.

{\itshape Proof of Lemma \ref{Adlm1}:}
Fixing arbitrary $x=(y,\Omega)\in \bbH^d(L)$ and $\lb\geq 0$,  the decomposition \eqref{ad4}  entails
\begin{eqnarray}\label{ad8}
	\sum_{\ell, m}|u_{\lb \ell m}(y,\Omega)|^2\is \sum_\ell N_{\lb \ell}(L)^2|q_{\lb \ell}(y)|^2\sum_{m}|Y_{\ell m}(\Omega)|^2
	\nonum
	\is \frac{1}{\text{vol}(\bbS^{d-1})} \sum_\ell G_\ell N_{\lb \ell}(L)^2|q_{\lb \ell}(y)|^2\,,
\end{eqnarray}
where the second line follows from Theorem \ref{thadd1} with $G_\ell = \frac{(2 \ell+d-2)(\ell+d-3)!}{\ell!(d-2)!}$. From Stirling's formula we obtain the asymptotic large $\ell$ behavior
\begin{eqnarray}\label{ad9}
	G_\ell N_{\lb \ell}(L)^{2}\sim 2^{-2\ell}\ell^{d-3}\,,
\end{eqnarray}
where the `$\sim $' means that $G_\ell N_{\lb \ell}(L)^{-2}/(2^{-2\ell}\ell^{d-3})$ is bounded by a ($\ell$-independent) constant as $\ell \to \infty$. 
Next, a  large $\ell$ bound on $|q_{\lb l}(y)|^2$ is obtained through the following integral representation of the hypergeometric function in \eqref{ad6},
\begin{eqnarray}
	F \big(\ell +a,\,\ell +b;\,\ell +c;-\sinh^2(y/2)\big)=\frac{\Gamma(\ell +c)}{\Gamma(\ell +b)\Gamma(c-b)}\int_0^\infty\!\!\! \frac{t^{\ell +b-1}(t+1)^{a-c}}{\big[\cosh^2(y/2)t+1\big]^{\ell +a}}\,\D t\,,\quad
\end{eqnarray}
where we have defined for brevity $a:=\rho +i\lb$, $b:=\rho-i\lb$, $c:=d/2$. Since for every $y \geq 0$ there is the upper bound $t^\ell/ [\cosh^2(y/2)\,t+1]^\ell \leq 1/(\cosh (y/2))^{2\ell}$, it follows that
\begin{align}\label{ad10}
	\big|F \big(\ell +a,\,&\ell +b;\,\ell +c;-\sinh^2(y/2)\big)\big|
	\nonum 
	&\leq \frac{|\Gamma(\ell +c)|}{|\Gamma(\ell +b)|\,|\Gamma(c-b)|}\frac{1}{(\cosh (y/2))^{2\ell}}\int_0^\infty\!\!\! \frac{t^{\rho-1}(t+1)^{-1/2}}{\big[\cosh^2(y/2)t+1\big]^{\rho}}\,\D t
	\nonum 
	&\leq \frac{1}{(\cosh (y/2))^{2\ell}}\, \ell^{1/2} K_\lambda \,\Big|F\Big(\rho, \,\rho;\,d/2;\,-\sinh^2(y/2)\Big)\Big|\,,
\end{align}
for all $\ell \in \bbN_0$ and $y\geq 0$, with $K_\lb $ an $\ell$-independent function of $\lb$. Together with Eq. \eqref{ad9} this gives the following large $\ell$-bound of the summands in \eqref{ad8}
\begin{align}\label{ad11}
	G_\ell N_{\lb \ell}(L)^2|q_{\lb \ell}(y)|^2\leq \ell^{d-2}\exp\big\{2\ell \ln[ \tanh(y/2)]\big\} K_\lambda \,\Big|F\Big(\rho, \,\rho;\,d/2;\,-\sinh^2(y/2)\Big)\Big|\,.
\end{align}
It is clear from \eqref{ad11} that the $\ell$-sum defining $d_\lb(x)$ converges uniformly over compact subsets of the form $[y_0,y_1]\times \bbS^{d-1}\subseteq \bbH^d(L)$, completing the proof of Lemma \ref{Adlm1}. \qed

\medskip

{\itshape Proof of Lemma \ref{Adlm2}:}
Fixing $\sigma\in SO(1,d)$ arbitrarily, we define for brevity $f_{\lb l}(x):=u_{\lb l}\circ \sigma(x)$, $x\in \bbH^d(L)$. Writing $x=(y,\Omega)$ for some arbitrary $y>0$, consider the $(d-1)$-sphere $\mcS$ (of radius $L\sinh y$) at this coordinate. It is in particular clear that $f_{\lb l}\big|_{\mcS}\in L^2(\bbS^{d-1})$, and hence for each  $y>0$ the function $f_{\lb l}$ may be expanded in spherical harmonics
\begin{eqnarray}\label{ad14}
	f_{\lb l}=\sum_{\jmath,m}\alpha_{\lb l\jmath m}(y)Y_{\jmath m}\,,\quad \alpha_{\lb l\jmath m}(y)=\int_{\bbS^{d-1}}\!\!\D\Omega\,Y_{\jmath m}^\ast f_{\lb l}\big|_{\mcS}\,,
\end{eqnarray}
where the convergence is in the $L^2(\bbS^{d-1})$.\footnote{By slight abuse of notation we have displayed the $y$-dependence of the coefficient functions $\alpha_{\lb j\ell m}$.} 
Unfortunately, convergence in $L^2$-norm is not sufficiently strong for our purposes. Instead, we now show that the sum in Eq. \eqref{ad14} converges {\itshape uniformly pointwise} on the (compact) set $[y_0,y_1]\times \bbS^{d-1}\subseteq \bbH^d(L)$, where $0<y_0<y_1$. To proceed, note that for $k\in \bbN$
\begin{eqnarray}\label{ad15}
	\big(\jmath(\jmath+d-2)\big)^k\alpha_{\lb l\jmath m}(y) \is \int_{\bbS^{d-1}}\!\!\D\Omega\, (-\Box_{\bbS^{d-1}})^k Y_{\jmath m}(\Omega)^\ast f_{\lb l}(y,\Omega)
	\nonum
	\is \int_{\bbS^{d-1}}\!\!\D\Omega\,  Y_{\jmath m}(\Omega)^\ast (-\Box_{\bbS^{d-1}})^k f_{\lb l}(y,\Omega)\,,
\end{eqnarray}
by repeated integration by parts on $\bbS^{d-1}$. In particular, $(-\Box_{\bbS^{d-1}})^k f_{\lb l}(y,\Omega)$ is a smooth function for all $y>0$, and its modulus is hence bounded by some $B_k>0$ on $[y_0,y_1]\times \bbS^{d-1}$ with $y_0>0$. Then
\begin{eqnarray}
	\big(\jmath(\jmath+d-2)\big)^k\big|\alpha_{\lb l\jmath m}(y)\big|\leq B_k\int_{\bbS^{d-1}}\!\!\D\Omega\,  |Y_{\jmath m}(\Omega)|\leq B_k\,\sqrt{\text{vol}(\bbS^{d-1})},
\end{eqnarray}
and hence
\begin{eqnarray}\label{ad16}
	\sup_{y\in [y_0,y_1]} |\alpha_{\lb l\jmath m}(y)\big|\leq \frac{B_k\,\sqrt{\text{vol}(\bbS^{d-1})}}{\big(\jmath(\jmath+d-2)\big)^k}\,.
\end{eqnarray}
For sufficiently large $k\in \bbN$ this implies the uniform pointwise convergence of 
\begin{eqnarray}\label{ad17}
	f_{\lb l}(y,\Omega)=\sum_{\jmath,m}\alpha_{\lb l\jmath m}(y)Y_{\jmath m}(\Omega)
\end{eqnarray}
over the compact subset $[y_0,y_1]\times \bbS^{d-1}\subseteq \bbH^d(L)$, proving the claim. Furthermore, this argument generalizes straightforwardly to show that 
\begin{eqnarray}\label{ad18}
	-\Box_{\bbH^d(L)}f_{\lb l}(y,\Omega)=\ssum{\jmath, m}{} -\Box_{\bbH^d(L)}\big(\alpha_{\lb l\jmath m}(y)Y_{\jmath m}(\Omega)\big)\,,
\end{eqnarray}
with the sum converging uniformly pointwise on $[y_0,y_1]\times \bbS^{d-1}\subseteq \bbH^d(L)$. Since $\sigma\in SO(1,d)$ is an isometry, its action commutes with that of the $-\Box_{\bbH^d(L)}$, so $-\Box_{\bbH^d(L)}f_{\lb l}=\mathscr{E}_\lb f_{\lb l}$. Therefore  $-\Box_{\bbH^d(L)}\big(\alpha_{\lb l\jmath m}(y)Y_{\jmath m}(\Omega)\big)=\mathscr{E}_\lb \alpha_{\lb l\jmath m}(y)Y_{\jmath m}(\Omega)$, and by separation of variables one concludes that $\alpha_{\lb l\jmath m}(y)\propto q_{\lb \jmath}(y)$, where we recall $u_{\lb \jmath m}(y,\Omega)=N_{\lb \jmath}q_{\lb \jmath}(y)Y_{\jmath m}(\Omega)$. In summary, the expansion \eqref{ad17} may be expressed in terms of the $u_{\lb j}$,
\begin{eqnarray}\label{ad19}
	u_{\lb l}\circ \sigma(y,\Omega)=\ssum{j}{}C_{\lb lj}(\sigma)u_{\lb j}(y,\Omega)\,,
\end{eqnarray}
where the summation is over the multi-index $j=(\jmath,m)$. This expression holds for all $y>0$, and the convergence is uniform over the compact sets $[y_0,y_1]\times \bbS^{d-1}$ with $y_0>0$. 

Next, since $\sigma$ is an isometry of $\bbH^d(L)$,  the functions  $u_{\lb l}\circ \sigma$ also constitute an orthonormal complete basis for $L^2(\bbH^d(L))$. The orthonormality and completeness of both sets $\{u_{\lb l}\circ \sigma\}$, $\{u_{\lb l}\}$ then clearly entails that the coefficients $C_{\lb jl}(\sigma)$ appearing in the expansion \eqref{ad19} must satisfy the unitarity relation $\ssum{k}{}C_{\lb k j}(\sigma)C_{\lb k l}(\sigma)^\ast =\delta_{jl}$ for all $\lb\in \bbR_{\geq 0}$.

Finally, the  large $l$ estimates on $u_{\lb l}$ developed in the proof of Lemma \ref{Adlm1}, together with the continuity of the functions $u_{\lb l}$ and $u_{\lb l}\circ \sigma$, implies that the result \eqref{ad19} can be continued to $x_0\in \bbH^d(L)$ defined by $y=0$. Thus  $u_{\lb l}\circ \sigma(x_0)=\ssum{j}{}C_{\lb lj}(\sigma)u_{\lb j}(x_0)$, with $u_{\lb j}(x_0)\neq 0$ for only finitely many $j$, as may be explicitly observed from Eqs. \eqref{ad4}, \eqref{ad6}.

%
%In order to obtain the result at $x_0\in \bbH^d(L)$ defined by $y=0$
%
%
%
%
%Having established the result \eqref{ad12} for all points on the 
%
%
%Finally, it follows from the large $l$ estimates on $u_{\lb l}$ developed in the proof of Lemma \ref{Adlm1}, together with the continuity of $u_{\lb l}$, $u_{\lb j}\circ \sigma$ that $u_{\lb j}\circ \sigma (y=0)=\ssum{l}{}C_{\lb jl}(\sigma)u_{\lb l}(y=0)$. Only finitely many terms in this sum are non-zero. \red{Small reminder: why?}

In conclusion, it has been shown  for any $\sigma\in SO(1,d)$, $x\in \bbH^d(L)$ there is pointwise convergence
\begin{eqnarray}
	(u_{\lb j}\circ \sigma)(x)\is \ssum{l}{}C_{\lb jl}(\sigma)u_{\lb l}(x)\,,\quad \ssum{k}{}C_{\lb jk}(\sigma)C_{\lb lk}(\sigma)^\ast =\delta_{jl}\,.
\end{eqnarray}
Furthermore, the convergence is uniform over all compact sets $[y_0,y_1]\times \bbS^{d-1}$ with $y_0>0$.
\qed

%%%%%%%%%%%%%%%%%%%
%	New section 
%%%%%%%%%%%%%%%%%%%

\section{Calculation of the gravitational effective action induced from a free scalar field on $\bbH^d(L)$}
\label{sec:PI}

Consider a free scalar field $A(x)$ on $\bbH^d(L)$ with action (c.f. \gl{D2})
\bg
S_\mrm{M}[A;g]=\foh\int\dd x\sg{x}\,A(x)\mK_x A(x) \, ,\quad \mK_x=-\Box_g^x+M^2+\xi R(x)\,.
\eg
This matter action induces the one-loop gravitational action $\Gamma_\mrm{1L}[g]$ via the path integral~\gl{D4},
\begin{equation}
\text{e}^{-\Gamma_\text{scal}[g]} = \int \mD\left[g^{1/4} A\right]\text{e}^{-S_\text{M}[A;g]} \equiv  \int \mD\left[B\right]\text{e}^{-S_\text{M}[g^{-1/4}B;g]}\;.
\end{equation}
The  scalar field density $B(x)$ may be expanded in the basis of densitized eigenfunctions $v_{\lambda l}^\text{S}:=g^{1/4}u_{\lambda l}^\text{S}$ of the Laplace operator as 
\begin{equation}\label{appC2}
B(x) = \int \text{d}\lambda \sum_l b_{\lambda l} v_{\lambda l}(x)\,,
\end{equation}
and the path-integral measure $\mathscr{D}B$ transforms formally as $\mathscr{D}B = \prod_{\lambda,l}\text{d}b_{\lambda l}$.\footnote{The Jacobian $\frac{\delta B(x)}{\delta b_{\lambda l}}= v_{\lambda l}(x) =: J_{x, \lambda,l}$ is one, since, by using the completeness and orthogonality relations,
$
	\left(JJ^\dagger\right)_{x,x'} = \int \text{d}\lambda \sum_l v_{\lambda l}(x)^\star v_{\lambda l}(x') = \delta(x-x')
	$ as well as
	$	\left(JJ^\dagger\right)_{\lambda l,\lambda'l'} = \int \text{d}^d x v_{\lambda l}(x)^\star v_{\lambda' l'}(x) = \delta(\lambda-\lambda')\delta_{ll'}$,
we have $\det J = \sqrt{\det\left(JJ^\dagger\right)}=1$.} 
Proceeding more carefully, since the  quantum number $\lambda$ is still continuous, it must be further discretized:
\begin{equation}
\lambda \equiv \lambda_n := n \varepsilon, \quad n \in \bbN_0, \quad \varepsilon \text{ small}\,.
\end{equation}
Then, the $\lambda$-integral in \eqref{appC2} becomes a sum of the form
\begin{equation}
\label{eq:C4dis}
\int \text{d}\lambda f(\lambda) =	 \varepsilon \sum_n f(\lambda_n)=: \varepsilon \sum_n f_n\;,
\end{equation}
and hence the measure reads
\begin{equation}
\mathscr{D}B = \prod_{\lambda,l}\text{d}b_{\lambda l} =\prod_{\lambda_n l}\text{d}b_{\lambda_n l}: = \prod_{n, l}\varepsilon \text{d}b_{n l}\;.
\end{equation}
Next, using the eigenvalue equation,
\begin{equation}
\mathscr{K}_x v_{\lambda l}^\text{S}(x) = \mathscr{F}_\lambda^\text{S}v_{\lambda l}^\text{S}(x), \quad \text{ where } \quad\mathscr{F}_\lambda^\text{S} = \mathscr{E}_\lambda^\text{S}+M^2 + \xi R \, ,
\end{equation}
with $ \mathscr{E}_\lambda^\text{S}$ given in Table \ref{table:eigenvalues}, the path integral becomes
\begin{eqnarray}
\text{e}^{-\Gamma_\text{scal}[g]} &=& \int \prod_{n, l} \varepsilon \text{d}b_{n, l} \exp \left\{-\frac{1}{2} \varepsilon^2 \sum_{n, n'} \sum_{l, l'}\mathscr{F}_n^\text{S}b_{nl}b_{n'l'} \;\int \text{d}^d x v_{nl}(x)v_{n'l'}(x)\right\}\nonumber\\&=&\int \prod_{n,l} \text{d}\left(\varepsilon b_{nl}\right) \exp\left\{-\frac{1}{2} \sum_{n,l}\mathscr{F}_n^\text{S}\left(\varepsilon b_{nl}\right)^2\right\} = \left(\prod_{n,l}\mathscr{F}_n^\text{S}/(2\pi)\right)^{-\frac{1}{2}}\;.
\end{eqnarray}
And hence we have, omitting irrelevant constant terms,\footnote{In the final step we use
$
		\delta(x) = \lim_{\varepsilon\to 0}\frac{1}{|\varepsilon| \sqrt{\pi}} \text{e}^{-\left(x/\varepsilon\right)^2} \ \Rightarrow\ \delta(0)\sim \lim_{\varepsilon \to 0} {1}/{|\varepsilon|}\nonumber\,.
$}
\begin{eqnarray}\label{appC1}
\Gamma_\text{scal}[g] &=&\frac{1}{2}\log \left(\prod_{n,l}\mathscr{F}_n^\text{S}\right)=\frac{1}{2}\sum_n \sum_l \log \mathscr{F}_n^\text{S}\nonumber\\
&=&\frac{1}{2}\frac{1}{\varepsilon}\varepsilon \sum_n \sum_l \log \mathscr{F}_n^\text{S}
\nonumber\\
&=& \frac{1}{2}\int \text{d}\lambda \delta(0) \lef(\sum_{l\in \mI(\mrm S)}1\ri) \log \mathscr{F}_\lambda^\text{S} \, ,
\end{eqnarray}
where, following Eq.~\gl{C4dis}, in the final step the discretization in $\lambda$ is removed to arrive at the (formal) result. Next, the expression in the final line of \eqref{appC1} is identified with the spectral function $\mu(\lambda)$ (c.f. \Gl{E16}) via the following formal computation (with $x_0$ referring to the `origin' of the hyperboloid)
\bg
\label{eq:E18d}
\spl{
\mu^\text{S}(\lambda)&:= \frac{\pi \Omega_{d-1}L^d}{2^{d-2}}\sum_{l\in\mI(\mrm{S})} \left(u^\text{S}_{\lambda l}\right)^* (x_0)\left(u^\text{S}_{\lambda l}\right) (x_0) \\
&=\frac{\pi \Omega_{d-1}L^d}{2^{d-2}}\frac{1}{\text{vol}\left[\bbH^d(L)\right]}\int \text{d}^dx\sqrt{g(x)} \sum_{l\in\mI(\mrm{S})}\left(u^\text{S}_{\lambda l}\right)^* (x_0)\left(u^\text{S}_{\lambda l}\right) (x_0)\\
&=\frac{\pi \Omega_{d-1}L^d}{2^{d-2}}\frac{1}{\text{vol}\left[\bbH^d(L)\right]}\int \text{d}^dx\sqrt{g(x)} \sum_{l\in\mI(\mrm{S})}\left(u^\text{S}_{\lambda l}\right)^* (x)\left(u^\text{S}_{\lambda l}\right) (x)
\\
&=\frac{\pi \Omega_{d-1}L^d}{2^{d-2}}\frac{1}{\text{vol}\left[\bbH^d(L)\right]}\sum_{l\in\mI(\mrm{S})}\int \text{d}^dx\sqrt{g(x)} \left(u^\text{S}_{\lambda l}\right)^* (x)\left(u^\text{S}_{\lambda l}\right) (x)
\\
&=\frac{\pi \Omega_{d-1}L^d}{2^{d-2}}\frac{1}{\text{vol}\left[\bbH^d(L)\right]}\,\delta(0)\!\sum_{l\in\mI(\mrm{S})} 1 \,.
}
\eg
To arrive at the second equality, one multiplies and divides by the volume of $\bbH^d(L)$, while the third line follows as Theorem \ref{thadd} entails constancy of the sum over $l\in\mI(\mrm{S})$. Finally, formally exchanging the order of the summation and integration, and exploiting the orthonormality of the eigenfunctions,
\begin{align}
	\int \text{d}^dx\sqrt{g(x)} \left(u^\text{S}_{\lambda l}\right)^* (x)\left(u^\text{S}_{\lambda' l'}\right) (x)=\delta(\lb-\lb')\delta_{ll'}\,,
\end{align}
yields the result.
Plugging this result into Eq. \eqref{appC1} yields, with $g_{\mu \nu} = L^2 \gamma_{\mu \nu}$,
\begin{eqnarray}
	\Gamma_\text{scal}[g] &\equiv& \Gamma_{\text{scal}}(L)=\frac{1}{2}\Tr_{\text{S}}\log (\mathscr{K})\nonumber\\&=& \frac{1}{2}\frac{2^{d-2}}{\pi \Omega_{d-1}} \frac{\text{vol}\left[\bbH^d(L)\right]}{L^d} \int \!\text{d}\lambda \,\mu^\text{S}(\lambda) \log \left(\mathscr{F}_\lambda^\text{S}(L)\right)\;.
\end{eqnarray}
Finally, we note that  an $N$-cutoff may be applied to the $\lb$-integration straightforwardly.

\newpage
\bibliography{references_2.bib}
\end{document}